\documentclass[english,emulateapj]{emulateapj}
\usepackage{graphicx}
\usepackage{amssymb}

\begin{document}

\title{VLA 1.4 GHz catalogs of the Abell 370 and Abell 2390 cluster fields }

\author{Isak G. B. Wold\altaffilmark{1}, Frazer N. Owen\altaffilmark{2},
Wei-Hao Wang\altaffilmark{3}, Amy J. Barger\altaffilmark{1,4,5},
and Ryan C. Keenan\altaffilmark{1}}

\altaffiltext{1}{Department of Astronomy, University of Wisconsin-Madison, 475 North Charter Street, Madison, WI 53706, USA; wold@astro.wisc.edu, barger@astro.wisc.edu}
\altaffiltext{2}{National Radio Astronomy Observatory, P.O. Box O, Socorro, NM 87801, USA; fowen@nrao.edu}
\altaffiltext{3}{Academia Sinica Institute of Astronomy and Astrophysics, P.O. Box 23-141, Taipei 10617, Taiwan}
\altaffiltext{4}{Department of Physics and Astronomy, University of Hawaii, 2505 Correa Road, Honolulu, HI 96822, USA}
\altaffiltext{5}{Institute for Astronomy, University of Hawaii, 2680 Woodlawn Drive, Honolulu, HI 96822, USA} 
\begin{abstract}
We present 1.4 GHz catalogs for the cluster fields Abell 370 and Abell
2390 observed with the Very Large Array. These are two of the deepest
radio images of cluster fields ever taken. The Abell 370 image covers
an area of 40$'$ $\times$ 40$'$ with a synthesized beam of $\sim$1.7$''$
and a noise level of $\sim$5.7 $\mu$Jy near field center. The Abell
2390 image covers an area of 34$'$ $\times$ 34$'$ with a synthesized
beam of $\sim$1.4$''$ and a noise level of $\sim$5.6 $\mu$Jy near
field center. We catalog 200 redshifts for the Abell 370 field. We
construct differential number counts for the central regions (radius
$<$ 16$'$) of both clusters. We find that the faint (S$_{1.4{\rm \: GHz}}<$
3 mJy) counts of Abell 370 are roughly consistent with the highest
blank field number counts, while the faint number counts of Abell
2390 are roughly consistent with the lowest blank field number counts.
Our analyses indicate that the number counts are primarily from field
radio galaxies. We suggest that the disagreement of our number counts
can be largely attributed to cosmic variance. 
\end{abstract}

\keywords{cosmology: observations}

\section{Introduction }

Radio surveys probe active galactic nuclei (AGNs) and star-forming
(SF) galaxies. SF galaxies produce non-thermal radio continuum through
synchrotron emission from supernova remnants. For these sources, the
1.4 GHz luminosity is found to be an accurate indicator of the star
formation rate \citep{condon92}. AGNs also emit synchrotron radiation,
which is ultimately powered by accretion onto supermassive black holes
(SMBHs). Observed tight correlations between SMBH mass and bulge properties
\citep{kormendy95,ferrarese00,tremaine02} plus various theoretical
considerations \citep{granato04,croton06,bower06} have led to the
hypothesis that AGNs regulate star formation. Thus, radio surveys
are an important tool to constrain how stars and SMBHs evolve and
interact as a function of cosmic time.

There is still debate over the faint radio population, especially
regarding the differential source counts. AGNs are found to dominate
the counts at high flux densities, while SF galaxies are thought to
emerge at lower flux densities \citep{condon89}. However, the exact
composition --- and the counts themselves --- are debated below S$_{1.4\:{\rm GHz}}$
$=$ 100 $\mu$Jy. The disagreement in faint radio counts from one
survey to the next has been attributed mostly to instrumental and
analysis effects, not to cosmic variance \citep{condon07}. For example,
for the heavily studied Hubble Deep Field-North (HDF-N), three different
groups have derived faint-end number counts, with each subsequent
study finding them to be incrementally higher \citep{richards00,biggs06,morrison10}.
However, studies such as \citet{biggs06}, which present catalogs
for three deep radio fields, have shown that similar instrument configurations
and analysis methods still result in faint counts that are inconsistent
with merely Poisson variation. It is important to obtain as many deep
radio images as possible to help resolve this issue, especially since
cosmic variance could be an important factor.

In this paper, we present deep radio observations of two heavily-studied
cluster fields. Radio surveys of cluster fields have been vital to
many areas of study. For example, they have improved our understanding
of cluster members by finding evidence for a population of dust-obscured
SF cluster galaxies that were previously classified as post-starburst
galaxies based on optical spectra. With the help of high-resolution
near-infrared (NIR) and optical imaging, \citet{smail99} interpreted
radio emission from these objects as an indication of ongoing star
formation. 

Radio surveys of cluster fields have also been useful in the study
of cluster evolution. For example, \citet[][]{morrison99} found that
the population of low-luminosity radio sources in clusters rapidly
increases with redshift (0.02 $\leq$ $z$ $\leq$ 0.41). This can
be interpreted as an extension of the Butcher-Oemler effect \citep[][]{butcher84},
since the majority of low-luminosity radio sources are found to be
blue SF galaxies \citep[][]{morrison03}. 

Radio data are key to multiwavelength studies. Radio emission is unobstructed
by dust, which avoids a major source of bias found in UV and optical
studies. With the addition of far-infrared (FIR) data, radio data
can be used to indicate the dominant emission mechanism. The radio
luminosity is observationally found to be tightly correlated with
the FIR power for SF galaxies locally \citep{helou85,condon92} and
at high redshift \citep{appleton04,ivison10,mao11}. Any significant
departure from the FIR-radio correlation is an indication of AGN activity. 

In addition, the positional accuracy of radio surveys can be used
to pinpoint counterparts at other wavebands. In the submillimeter,
single dish telescopes have very poor resolution, and the unambiguous
identification of counterparts is not possible without additional
information. From the FIR-radio correlation, we know that FIR luminous
SF galaxies, such as submillimeter galaxies (SMGs), are correspondingly
luminous in radio emission. Thus, radio data can be used to identify
SMG counterparts. \citet{barger00}, \citet{ivison02}, and \citet{chapman03}
found that $\sim$60\% of bright ($>$2 mJy) SMGs had radio counterparts.
This feature allowed \citet{chapman05} to use radio positions to
target spectroscopically a large sample of bright SMGs, establishing
the redshift distribution for this population. Bright SMGs are predominantly
massive, dust-obscured, SF galaxies at a median redshift of $\sim$2.2
\citep{chapman05,alexander05}.

Submillimeter observations are a redshift independent probe (due
to a negative $K$-correction: 1 $<$ $z$ $<$ 8) of dust-reprocessed
UV light. However, the positive $K$-correction of the radio synchrotron
emission results in faint observed 1.4 GHz fluxes for high-redshift
objects. Thus, a high submillimeter-to-radio flux measurement is an
indication of a high-redshift source (Wang et al.\ 2007, 2009; \citealt{danner08};
Capak et al.\ 2008, 2011; Schinnerer et al.\ 2008; Daddi et al.\ 2009a,b; Coppin et al.\ 2009, 2010; Riechers et al.\ 2010;
\citealt{knudsen10}). This extreme population of very massive sources
at very high redshifts has been the topic of intense study and has
led many authors to suggest the need for significant modifications
to models of galaxy evolution \citep[e.g., ][]{granato04,baugh05,hopkins05,lukic07}.
Deep radio surveys continue to be an important tool in the discovery
of these objects.

Blank-field submillimeter surveys with the Submillimeter Common-User
Bolometer Array \citep[SCUBA; ][]{holland99} first resolved the bright
SMGs that account for $\sim20\%-30\%$ of the 850 $\mu$m extragalactic
background light \citep[e.g., ][]{barger98,hughes98,barger99,eales99}.
These surveys cannot reach the sensitivities required to detect directly
the dominant population of $<2$ mJy sources because of confusion
noise resulting from the coarse resolution of SCUBA. In order to detect
this population, one must observe fields with massive cluster lenses
to take advantage of both gravitational amplification by the lens
and reduced confusion noise \citep[][]{smail97,cowie02,knudsen08}.
Deep radio images of cluster fields are therefore valuable for identifying
the counterparts and determining the properties of this important
faint SMG population. Such data will also be very useful for helping
to interpret \textit{Herschel} \citep[][]{egami10} and SCUBA2 (C.-C. Chen et al.\ 2012,
in preparation) observations. 

In this paper, we construct deep radio catalogs of the cluster fields
Abell 370 (A370) and Abell 2390 (A2390), which we have reduced and
analyzed in a similar fashion. We derive the number counts for the
two fields, and we discuss the influence of the cluster and the importance
of cosmic variance on these counts. In our A370 catalog we include
redshifts for 200 radio sources that we obtained from the literature,
from unpublished work, and from our own observations. We note that,
in addition to copious ancillary data, these clusters also have excellent
lens models \citep[][]{kneib93,kneib02,richard10}. Our radio catalogs
in combination with the public data already available for these extensively
studied fields will aid in both cluster and field galaxy studies.
\begin{figure*}[!t]
\begin{raggedright}
\includegraphics[bb=12bp 140bp 1000bp 1100bp,clip,scale=0.27]{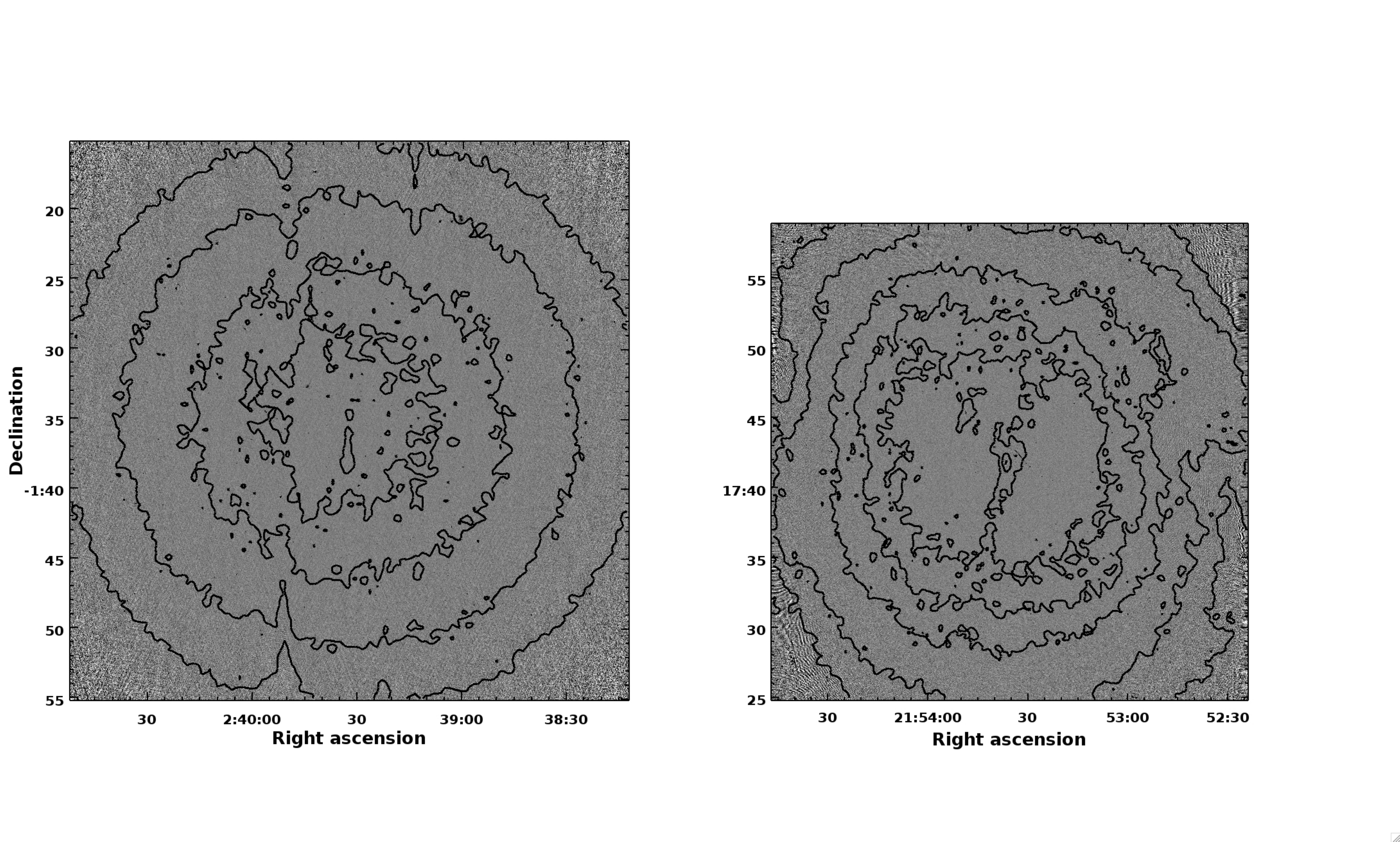}\includegraphics[bb=1127bp 140bp 1950bp 970bp,clip,scale=0.27]{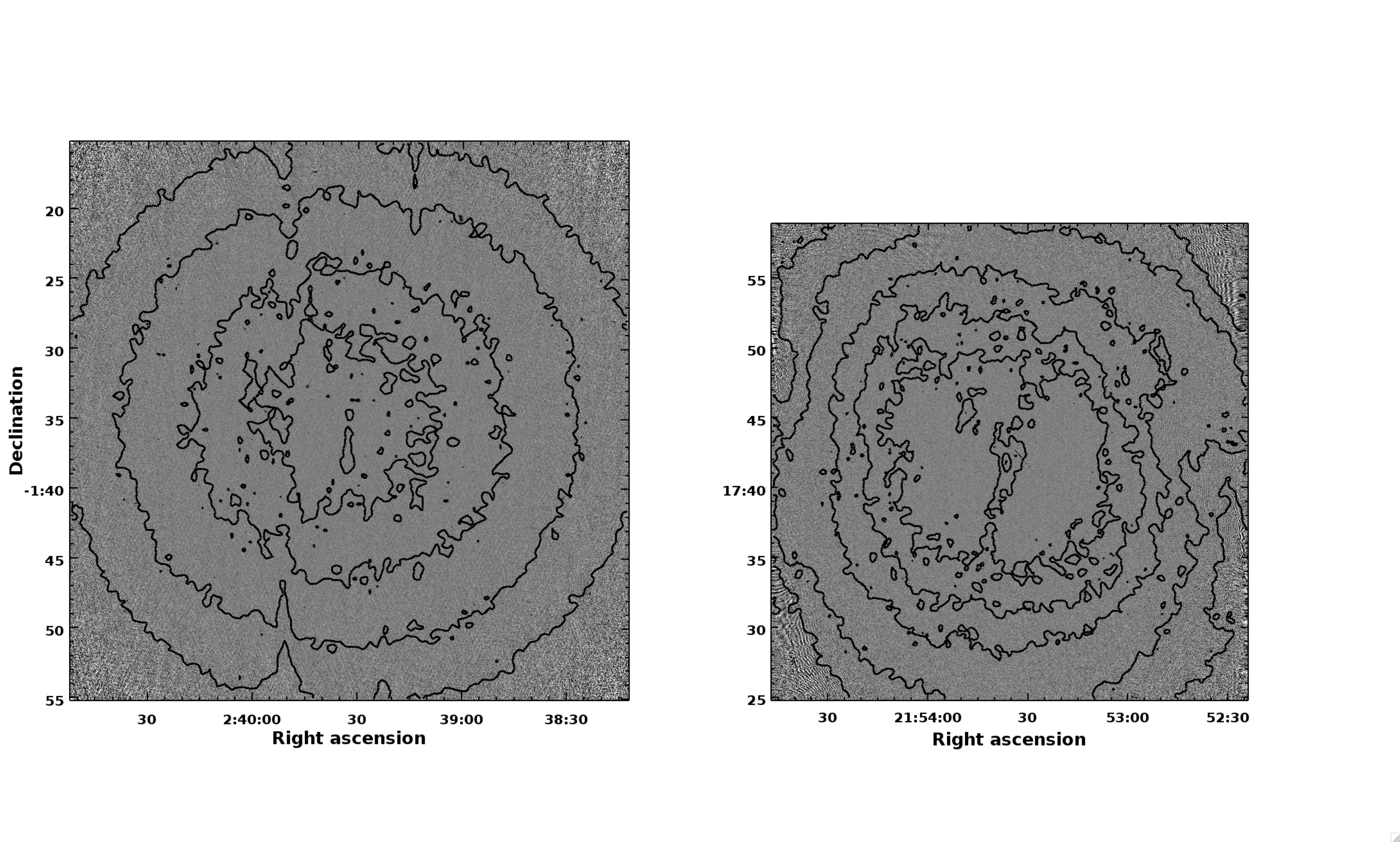}
\par\end{raggedright}

\caption{\textbf{\textit{Left}}: Contours of constant rms noise overlaid on
the 40$'$$\times$40$'$ A370 image. Contours levels are 6.5, 8.0,
12.0, and 20.0 $\mu$Jy and are located approximately 6$'$, 11$'$,
16$'$, and 20$'$ from field center. The image has a 1$\sigma$ rms
noise level of $\sim$5.7 $\mu$Jy near field center and a 1.8$''$
$\times$ 1.6$''$ synthesized beam.\textbf{\textit{ Right}}: Contours
of constant rms noise overlaid on the 34$'$$\times$34$'$ A2390
image. Contours levels are 5.5, 6.5, 8.0, 12.0, and 20.0 $\mu$Jy
and are located approximately 8$'$, 10$'$, 12$'$, 16$'$, and 19$'$
from field center. The image has a 1$\sigma$ rms noise level of $\sim$5.6
$\mu$Jy near field center and a 1.4$''$ $\times$ 1.3$''$ synthesized
beam. }
\end{figure*}
\begin{figure*}[!t]
\begin{centering}
\includegraphics[bb=15bp 9bp 1550bp 1524bp,clip,scale=0.3]{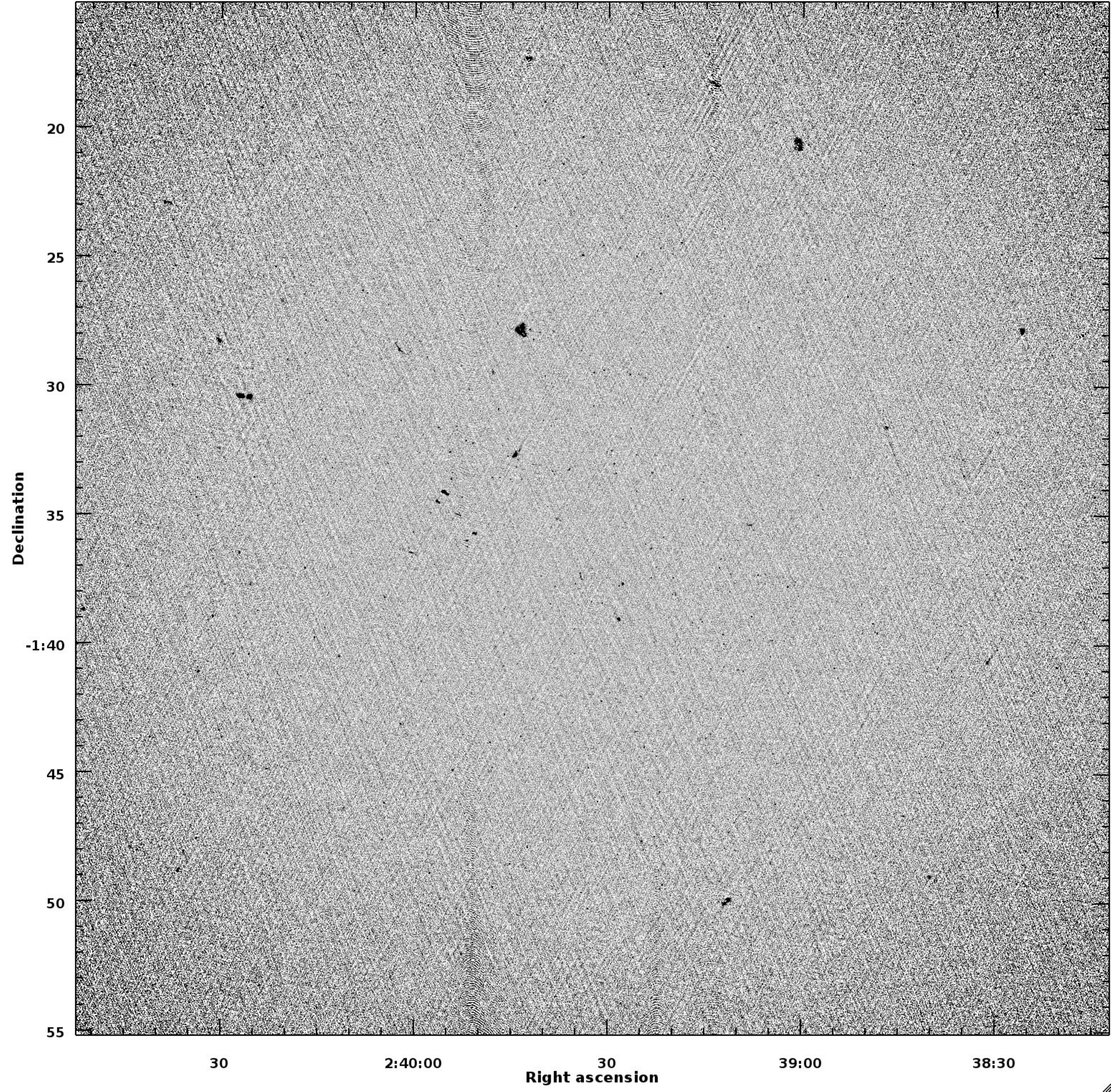}
\par\end{centering}

\caption{The 40$'$$\times$40$'$ A370 image. Field center is located at $02^{h}39^{m}32^{s}$,
$-01^{\circ}35'07''$ in J2000 coordinates. This is offset by $\sim$5$'$
from the cluster center at $02^{h}39^{m}50.5^{s}$, $-01^{\circ}35'08''$.
We observe no radio halos or relics; however, smaller array configurations
would be more sensitive to these large structures.}
\end{figure*}
\begin{figure*}[!t]
\begin{centering}
\includegraphics[bb=15bp 10bp 1550bp 1524bp,clip,scale=0.3]{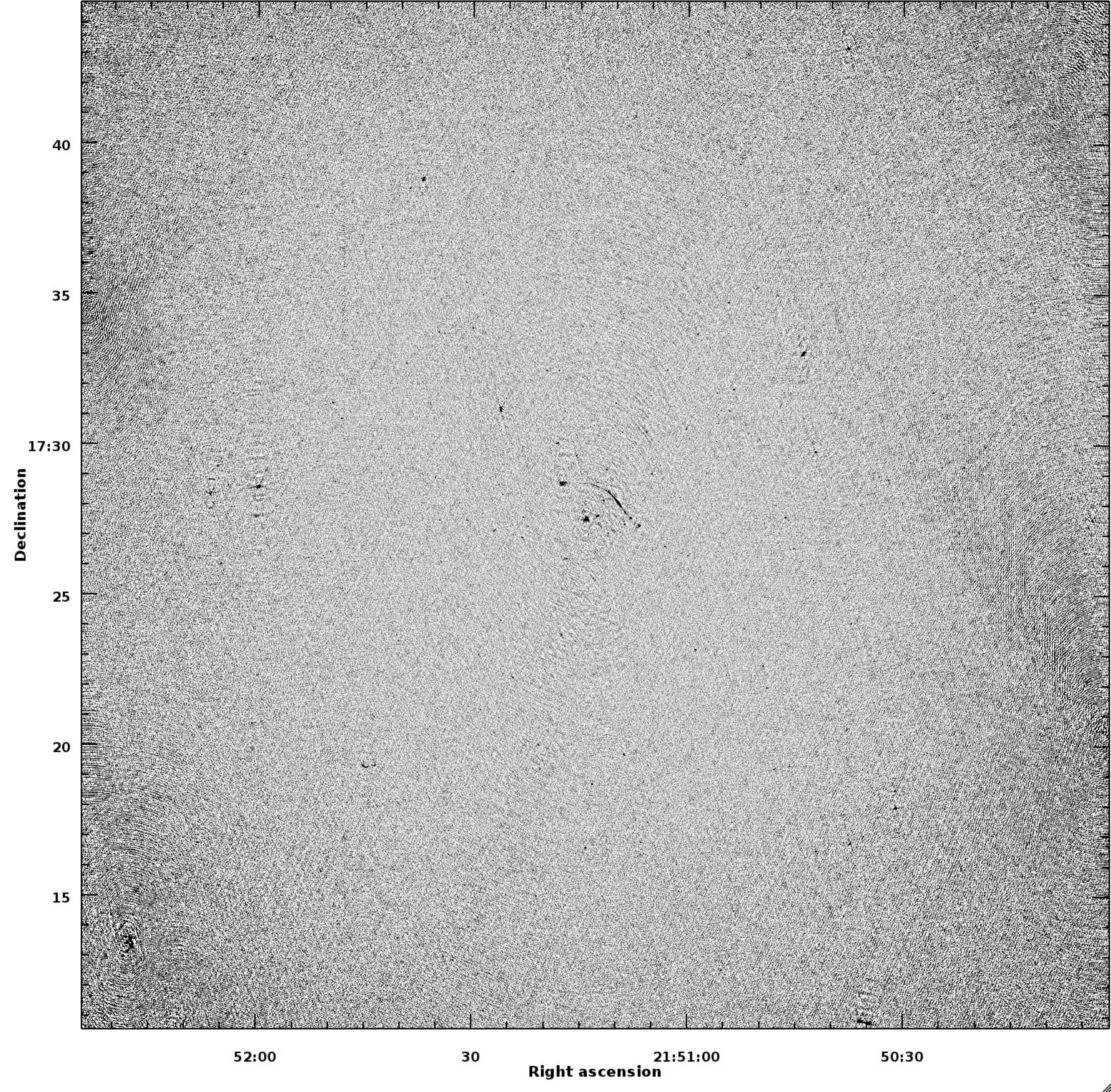}
\par\end{centering}

\caption{The 34$'$$\times$34$'$ A2390 image. Field center is located at
$21^{h}53^{m}36^{s}$, $+17^{\circ}41'52''$ in J2000 coordinates.
This is only offset by 20$''$ from the cluster's central cD galaxy
at $21^{h}53^{m}37^{s}$, $+17^{\circ}41'44''$. We observe no radio
halos or relics; however, smaller array configurations would be more
sensitive to these large structures.}
\end{figure*}

We describe our observations and data reduction in Section 2. In Section
3, we describe our source extraction and cataloging. In Section 4,
we compare our catalog to large-area survey results for bright sources.
In Section 5, we construct differential number counts for the central
regions (radius $<$ 16$'$) of both clusters. We discuss the derived
counts in Section 6 and summarize the paper in Section 7. Throughout
this paper, our adopted cosmology is H$_{0}$ $=$ 70 km s$^{-1}$
Mpc$^{-1}$, $\Omega_{M}$ $=$ 0.3, $\Omega_{\Lambda}$ $=$ 0.7.

\section{OBSERVATIONS AND DATA REDUCTION}

\subsection{A370 VLA Observations and Data Reduction}

We observed the A370 cluster field with the VLA in the A configuration
for $\sim$42.4 hr on-source during August and September 1999. K.
S. Dwarakanath observed A370 in the B configuration for $\sim$18.4
hr on-source during August and September 1994. In Table 1, we summarize
the parameters of the observing runs. In both configurations, the
data were taken in spectral line mode 4, which records seven 3.125
MHz channels in each of the two intermediate frequencies (IFs) centered
at 1.365 and 1.435 GHz, and each of the two polarizations. Five and
ten second integration times were utilized for the A and B configurations,
respectively. Field center is located at $02^{h}39^{m}32^{s}$, $-01^{\circ}35'07''$
in J2000 coordinates. This is offset by approximately 5 arcmins from
the cluster center at $02^{h}39^{m}50.5^{s}$, $-01^{\circ}35'08''$.
\begin{table}
\begin{centering}
\caption{Observing Summary\vspace{-0.5cm}
}

\par\end{centering}

\centering{}%
\begin{longtable}{ccccc}
\hline
\hline 
Field & Config. & Start Date & End Date & On Source Hrs\tabularnewline
\endhead
\hline 
A370 & A & 1999 Aug 24 & 1999 Sep 20 & 42.4\tabularnewline
A370 & B & 1994 Aug 01 & 1994 Sep 03 & 18.4\tabularnewline
A2390 & A & 2008 Oct 20 & 2008 Oct 28 & 31.4\tabularnewline
\hline 
\end{longtable}
\end{table}
\addtocounter{table}{-1}

Since we reduced both the A and B configuration data in a similar
fashion, we only describe our reduction of the B configuration data.
When necessary, we indicate any significant alterations needed for
the A configuration data reduction. Once we completed our initial
reduction, we combined the two configuration data sets and performed
a final round of calibration. Our data reduction techniques follow
the procedures described by \citet[][hereafter OM08]{owen08} and
\citet{owen09}. We exclusively used the AIPS package to reduce these
data.

We reduced the data obtained during the first day of the observations
in the following manner. We edited the phase and flux calibrators
for obvious amplitude anomalies using TVFLG. Then, we split the flux
calibrator from the raw database and applied a phase self-calibration.
At this point, we determined the bandpass correction using the BPASS
task applied to the flux calibrator. To account for the effects of
the sloping spectral response, we applied this bandpass correction
by copying the table generated by BPASS to the raw database. We then
scaled the primary flux calibrator, 3C 48, using the Baars flux density
values \citep{baars77}. 

We imaged these data using the AIPS task IMAGR and the three dimensional
multifacet options. To image $0.9^{\circ}$ from field center and
include all bright NVSS objects within a $2.5^{\circ}$ radius required
211 facets, each with 512$^{2}$ 1.5$''$ pixels. The A configuration
required 384 facets (each with 1024$^{2}$ $0.5''$ pixels) to image
$0.8^{\circ}$ from field center and include all strong B configuration
sources. After initial imaging, we limited cleaning to real features
by designating detected sources with clean boxes. We performed phase
self-calibration followed by phase and amplitude self-calibration.
Thus, we generated clean images which were then used as fiducial models
for each subsequent day.

We employed the VLA pipeline as recommended by AIPS Memo 112 (Lorant
Sjouwerman 2007) to perform initial calibration on subsequent days.
Then, we calibrated these data to images produced from the day one
observations. We used the AIPS procedure STUFFR to combine and compress
data from all days of the observing run, producing one master visibility
file (for each configuration).

We then split this combined visibility data set by IF and further
self-calibrated. As discussed in \citet{owen09}, we used UVSUB to
subtract the bright off-axis sources from the visibility data. We
then imaged the facets known to contain bright, outlying sources as
a data cube. We subtracted the generated clean components from the
original visibility data channel by channel. We conducted this residual
spectral subtraction procedure to compensate for frequency-dependent
artifacts which produce radial smearing for very bright sources far
from field center.

We then combined the like components of the A and B array with DBCON
and further split the data into 2 polarizations, thus producing 4
visibility files (2 polarizations and 2 IFs). We further calibrated
these data, including another iteration of the residual spectral subtraction
procedure. As a final calibration step, we employed PEELR to account
for local gain variations produced by bright sources in the outskirts
of the field. Final images were made using robust weighting (ROBUST=0
in IMAGR) and using IMAGR's UVTAPER empirically set to 100,100. We
combined the facet images into 4 composite images using FLATN. Then,
we combined these 4 images, weighted by rms$^{-2}$. Finally, we flattened
the image composite using MWF (as performed in OM08, but with a 41
$\times$ 41 pixel support window). This final image has a noise level
of $\sim$5.7 $\mu$Jy rms near field center, with a $1.8''\times1.6''$
synthesized beam at a position angle of $166^{\circ}$. In Figure
1(a), we show the 40$\times$40 arcmin$^{2}$ A370 radio field with
rms noise contours overlaid. Residual sidelobes from bright outlying
sources are still clearly present on either side of the field center,
offset by approximately 6$'$ from center with a North-South orientation.
In Figure 2, we show a larger image of the A370 radio field without
noise contours to allow for inspection of individual radio sources.
We observe no radio halos or relics; however, smaller array configurations
would be more sensitive to these large structures.

\subsection{A2390 VLA Observations and Data Reduction}

We observed the A2390 cluster field with the VLA in the A configuration
for $\sim$31.4 hr on-source during October 2008. We obtained the
data in spectral line mode 4, and we sampled the data every 3.3 seconds.
Field center is located at $21^{h}53^{m}36^{s}$, $+17^{\circ}41'52''$
in J2000 coordinates.

As described above for the A370 field, we followed the data reduction
techniques outlined by OM08 and \citet{owen09}. The primary flux
calibrator was 3C 48. We used 102 facets to cover the primary beam.
Each facet consisted of 1024$^{2}$ 0.5$''$ pixels. We carried out
the reductions as described above, producing a master visibility file
for all A configuration data. After an initial pass of self calibration,
we split the visibility file into the 28 parts that come from the
7 channels, 2 IFs, and 2 polarizations. We self-calibrated, imaged,
cleaned, and PEELRed the 28 data sets independently. We carried out
a few self-calibration/imaging/PEELR cycles to obtain a converging
solution for each of the 28 data sets. We then made 28 images with
the AIPS task IMAGR. This procedure, which further splits the visibility
files by channel, was necessary to minimize the sidelobes from the
bright sources around the edges of the field and from the central
cD galaxy.

We combined the 28 images using IDL. For each pixel in the 28 images,
we first measured the local background fluctuation in a 32$''$ $\times$
32$''$ region around it. Such background fluctuations come from both
noise and weak residual sidelobes from bright sources that cannot
be perfectly calibrated with the above procedure. We then merged the
28 images using an error-weighted mean according to the above measured
background fluctuation. Very weak residual sidelobes can still be
seen in the final merged image in certain regions, especially along
the North-South side of the central cD galaxy. However, we obtained
an excellent overall 5.6 $\mu$Jy rms in the central part of the image.
The synthesized beam is $1.4''\times1.3''$ with a position angle
of $72^{\circ}$. In Figure 1(b), we show the 34$\times$34 arcmin$^{2}$
A2390 radio field with rms noise contours overlaid. In Figure 3, we
show a larger image of the A2390 radio field without noise contours
to allow for inspection of individual radio sources. We observe no
radio halos or relics; however, smaller array configurations would
be more sensitive to these large structures.

\begin{deluxetable*}{ccccccccccccc}
\tablewidth{0pc}
\tablecaption{Sample of Radio Source Catalog}
\tablehead{
\colhead{Abell} & \colhead{Number} & \colhead{R.A. (J2000)} & \colhead{Dec.} & \colhead{PNR} & \colhead{S$_{1.4}~\pm$ e} & \colhead{} & \colhead{Size} & \colhead{} & \colhead{Upper} & \colhead{Beam} & \colhead{Redshift} & \colhead{Ref.}\\
\colhead{Field} & \colhead{} & \colhead{hms $\pm$ e(s)} & \colhead{d $'$ $''$ $\pm$ e($''$)} & \colhead{} & \colhead{$\mu$Jy} & \colhead{Maj} & \colhead{Min} & \colhead{P.A.} & \colhead{$''$} & \colhead{$''$} & \colhead{} & \colhead{}\\
\colhead{(1)} & \colhead{(2)} & \colhead{(3)} & \colhead{(4)} & \colhead{(5)} & \colhead{(6)} & \colhead{(7)}  & \colhead{(8)}  & \colhead{(9)} & \colhead{(10)} & \colhead{(11)} & \colhead{(12)} & \colhead{(13)}}
\startdata
     370 &        0 & 02 38 13.15 0.439 & -01 46 26.57 0.55 &  25.4 &  1422.9$\pm$ 74.3 & 5.8 & 3.9 &  59 & \nodata & 1.8 & \nodata & \nodata\\
     370 &        1 & 02 38 13.83 0.095 & -01 17 30.36 0.12 &   5.3 &   705.1$\pm$134.5 & \nodata & \nodata & \nodata & 0.5 & 3.0 & \nodata & \nodata\\
     370 &        2 & 02 38 14.64 0.070 & -01 34 15.05 0.06 &  23.5 &   610.7$\pm$ 31.8 & \nodata & \nodata & \nodata & 1.6 & 1.8 & \nodata & \nodata\\
     370 &        3 & 02 38 15.07 0.192 & -01 51 24.19 0.22 &   7.9 &   575.6$\pm$ 75.1 & \nodata & \nodata & \nodata & 3.2 & 3.0 & \nodata & \nodata\\
     370 &        4 & 02 38 15.50 0.110 & -01 55  2.76 0.10 &   5.3 &  1005.1$\pm$190.3 & \nodata & \nodata & \nodata & 1.8 & 1.8 & \nodata & \nodata\\
     370 &        5 & 02 38 16.61 0.292 & -01 33  0.89 0.21 &  13.0 &   207.8$\pm$ 25.6 & 7.2 & 0.0 &  60 & \nodata & 1.8 & \nodata & \nodata\\
     370 &        6 & 02 38 16.62 0.400 & -01 27 58.24 0.53 & 934.8 & 31519.4$\pm$946.3 & 6.7 & 3.8 & 103 & \nodata & 1.8 & \nodata & \nodata\\
     370 &        7 & 02 38 16.89 0.416 & -01 53  6.26 0.54 &  39.4 &  4367.0$\pm$177.2 & 6.7 & 3.9 &  42 & \nodata & 1.8 & \nodata & \nodata\\
     370 &        8 & 02 38 17.26 0.421 & -01 50 41.28 0.54 &  32.2 &  3195.5$\pm$145.4 & 6.0 & 3.2 &  48 & \nodata & 1.8 & \nodata & \nodata\\
     370 &        9 & 02 38 17.35 0.221 & -01 52 46.70 0.37 &   5.3 &   390.1$\pm$ 73.9 & \nodata & \nodata & \nodata & 4.2 & 1.8 & \nodata & \nodata\\
\enddata
\tablecomments{Table 2 is published in its entirety in the electronic edition of \textit{Astrophysical Journal Supplement}. A portion is shown here for guidance regarding its form and content.}
\end{deluxetable*}

\subsection{A370 Spectroscopic Observations and Reductions}

We targeted 58 radio sources without existing spectral data using
the Hydra fiber spectrograph on the Wisconsin-Indiana-Yale-NOAO (WIYN)
telescope. We preferentially targeted optically bright galaxies. We
observed these data with a single two hour pointing on 2012 January
20. Seeing was approximately 0.8$''$. We configured the spectrograph
using the ``red'' fiber bundle and the 316@7.0 grating at first
order with the GG-420 filter to provide a spectral window of $\sim$
4500 $-$ 9500 \AA ~with a pixel scale of 2.6 \AA ~per pixel.
The Hydra ``red'' fibers are 2$''$ in diameter and have a positional
accuracy of 0.3$''$, which ensured that the majority of light from
our target galaxies was observed with little contamination from the
sky and neighboring sources. We employed the IRAF task \textit{dohydra}%
\footnote{http://iraf.noao.edu/tutorials/dohydra/dohydra.html%
} in the reduction of our spectra. This task is specifically designed
for reduction of data from the Hydra spectrograph and includes steps
for dark and bias subtraction, flatfielding, dispersion calibration,
and sky subtraction. As described in Keenan et al.\ (2012), we used
the IRAF task \textit{xcsao} within the \textit{rvsao} package \citep{kurtz98}
to determine redshifts for our observed spectra. Of the 58 targets,
we obtained high-confidence redshifts for 36.

\section{Source extraction and cataloging}

For the A370 image, we extracted sources from the central 40$\times$40
arcmin$^{2}$ region. For the A2390 image, we extracted sources from
the central 34$\times$34 arcmin$^{2}$ region. To help account for
instrumental and intrinsic image broadening effects, we also analyzed
our images at lower resolutions. For each field, we created two complementary
low resolution images by convolving the final images with Gaussians
with FWHM equal to 3$''$ and 6$''$. We extracted sources from the
full, 3$''$, and 6$''$ resolution images independently. We then
compared and collated the results. This procedure has been shown to
assist in the detection of low surface brightness sources \citep{morrison10}.

We generated noise maps using the AIPS routine RMSD (IMSIZE=71,-1;
OPTYPE='HIST') for each of the three resolutions. When extracting
sources using the AIPS routine SAD, we used the output of RMSD to
estimate the background noise. SAD searches an image for sources above
some peak flux threshold and then uses a Gaussian fitting routine
to estimate flux density. For each resolution, we extracted sources
with peak signal-to-noise ratios (PNRs) greater than 4. For sources
with PNR $\leq$ 5.5 or for residual sources found manually, we used
JMFIT to determine source properties. The final catalog contains sources
with PNRs greater than 5. We accounted for bandwidth smearing and
primary beam attenuation in both the SAD and JMFIT procedures.

We considered a source resolved if the lower limit for the major axis
is greater than zero and the integrated flux minus 1$\sigma$ error
exceeds the peak flux. For unresolved sources, we recorded the upper
limit of the major axis and the total flux (best estimated by the
peak flux, OM08). For resolved sources, we recorded the integrated
flux and the best fit deconvolved major, minor, and position angle
(P.A.). 

For extended sources, it is more accurate to estimate the flux densities
using TVSTAT, which sums the total signal in a user-defined area,
rather than using the summation of the SAD Gaussians. As prescribed
by \citet{morrison10}, we used IMEAN to determine the peak flux and
its location. Using the noise map, we then measured the 1$\sigma$
noise level at this position. We adopted the ratio of these values,
peak flux / 1$\sigma$ noise level, as the PNRs for extended sources.

For each field, we compared the three catalogs extracted from the
three images with different resolutions. We adopted the 3$''$ resolution
source only if the PNR was more than 10\% higher than its higher resolution
counterpart. Similarly, we adopted the 6$''$ resolution source only
if the PNR was more than 10\% higher than its 3$''$ resolution counterpart.
Thus, we generated a catalog for the A370 and A2390 fields. 

We compared an intermediate image constructed just after combination
of the A and B configurations to our final A370 image. We found that
the extracted source flux is underestimated by a factor of 1.1. We
attribute this discrepancy to calibration errors from combining the
A and B configuration data. In our data reduction routine, we performed
self calibration of the combined data without constraining the mean
gain modulus to unity. In the presented A370 catalog, we have corrected
for this systematic offset. 

We present the first ten lines of our combined A370 and A2390 catalog
in Table 2. In Column 1, we indicate the Abell field, either Abell
370 or 2390. In Column 2, we indicate the radio source number. There
are a total of 699 galaxies in the A370 catalog and 524 galaxies in
the A2390 catalog. In Columns 3 and 4, we give the (J2000) right ascension
and declination with error estimates. We report the peak signal-to-noise
ratio in Column 5. In Column 6, we indicate the total 1.4 GHz flux
and uncertainty in $\mu$Jy. For resolved sources, we list the best-fit
deconvolved size characteristics in Columns 7$-$9, reporting the
major and minor FWHM sizes in arcsec. For unresolved sources, we indicate
the major axis upper limit size in Column 10. We list the resolution
of the image used in the source extraction in Column 11. 

Only the A370 field contains data in Columns 11 and 12. In these columns,
we list 200 A370 redshifts and their references. To identify redshift
- radio matches, we visually inspected all spectroscopic data lying
within 10$''$ of a radio source position by overlaying radio contours
and the spectral position on deep broadband optical images. We adopted
this technique, because extended and double-lobe radio sources often
have peak radio brightness offset from their optical center. Although
these redshifts have been utilized in previous studies, the vast majority
are published for the first time. Reference 1 designates redshifts
obtained via an ongoing spectroscopic campaign using the DEep Imaging
Multi-object Spectrograph (DEIMOS) on the Keck telescope (L. Cowie,
private communication). Reference 2 denotes a campaign by F. Owen
using Hydra on WIYN (F. Owen, private communication). Reference 3
designates additional WIYN Hydra redshifts obtained in a $z$ $\sim$
0.2 NIR follow-up survey (Keenan et al.\ 2012). Reference 4 denotes
an archival redshift obtained from \citet{dunlop89}. Reference 5
designates the new WIYN Hydra redshifts that we obtained (see Section
2.3) to supplement the existing data. We determine the redshift false
identification rate by shifting the radio image by 70$''$ and re-identifying
redshift - radio matches. We estimate that 4$\%$ of our redshift
identifications are false.

\begin{figure}
\begin{centering}
\includegraphics[bb=160bp 80bp 690bp 540bp,clip,angle=180,scale=0.4]{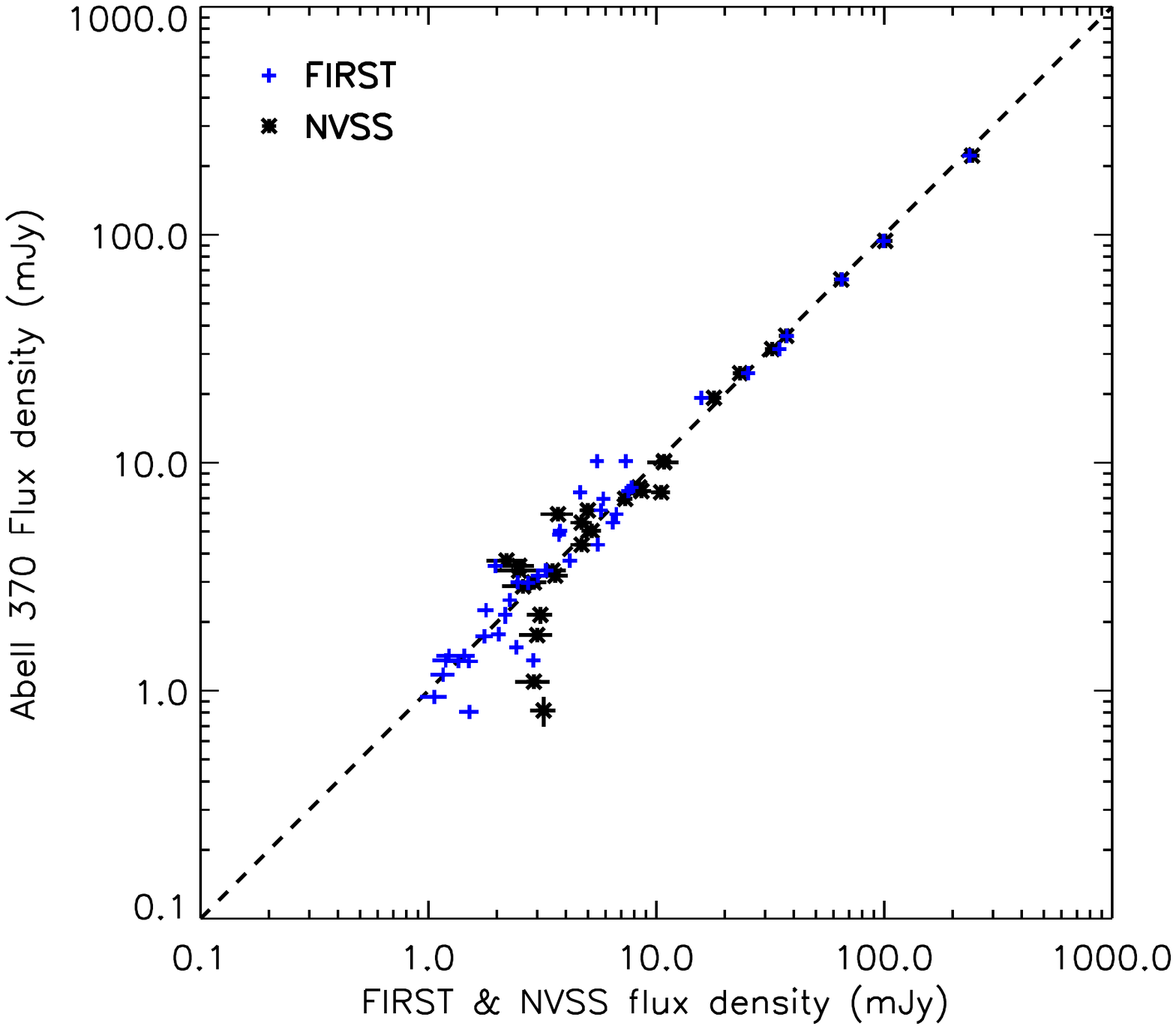}\caption{A370 flux densities compared to those from the FIRST and NVSS catalog.
41 out of 50 FIRST objects (blue crosses) and 29 out of 30 NVSS objects
(black asterisks) are found to have counterparts. The dashed line
indicates the ideal case, equality of flux densities. }

\par\end{centering}

{\footnotesize (A color version of this figure is available in the
online journal.)}
\end{figure}

\section{Comparison with NVSS and FIRST}

We compare our work to the Faint Image of the Radio Sky at Twenty-cm
(FIRST) \citep{white97} and the NRAO VLA Sky Survey (NVSS) \citep{condon98}
catalogs. The FIRST survey is a VLA program which produced images
with 0.75 mJy sensitivity and a resolution of $\sim$5$''$ in the
vicinity of A370. This survey has 50 sources coincident with our 40$\times$40
arcmin$^{2}$ A370 field. Of these 50 objects, 41 have counterparts
in our catalog. NVSS provides $\sim$2.5 mJy sensitivity and a resolution
of 45$''$. There are 30 NVSS sources in our A370 field. Twenty-nine
of these have counterparts in our radio catalog. 

Based on all 41 FIRST sources with counterparts in our radio catalog,
the mean($S_{1.4\:{\rm GHz}}^{{\rm FIRST}}$/$S_{1.4\:{\rm GHz}}^{{\rm A370}}$)
= 1.01$\pm$0.05. Based on the 25 sources brighter than the $\sim$2.5
mJy NVSS sensitivity limit that also have counterparts, the mean($S_{1.4{\rm \: GHz}}^{{\rm NVSS}}$/$S_{1.4\:{\rm GHz}}^{{\rm A370}}$)
= 0.98$\pm$0.04. Figure 4 compares our flux density measurements
with those of NVSS and FIRST.
\begin{figure}
\centering{}\includegraphics[bb=160bp 80bp 690bp 540bp,clip,angle=180,scale=0.4]{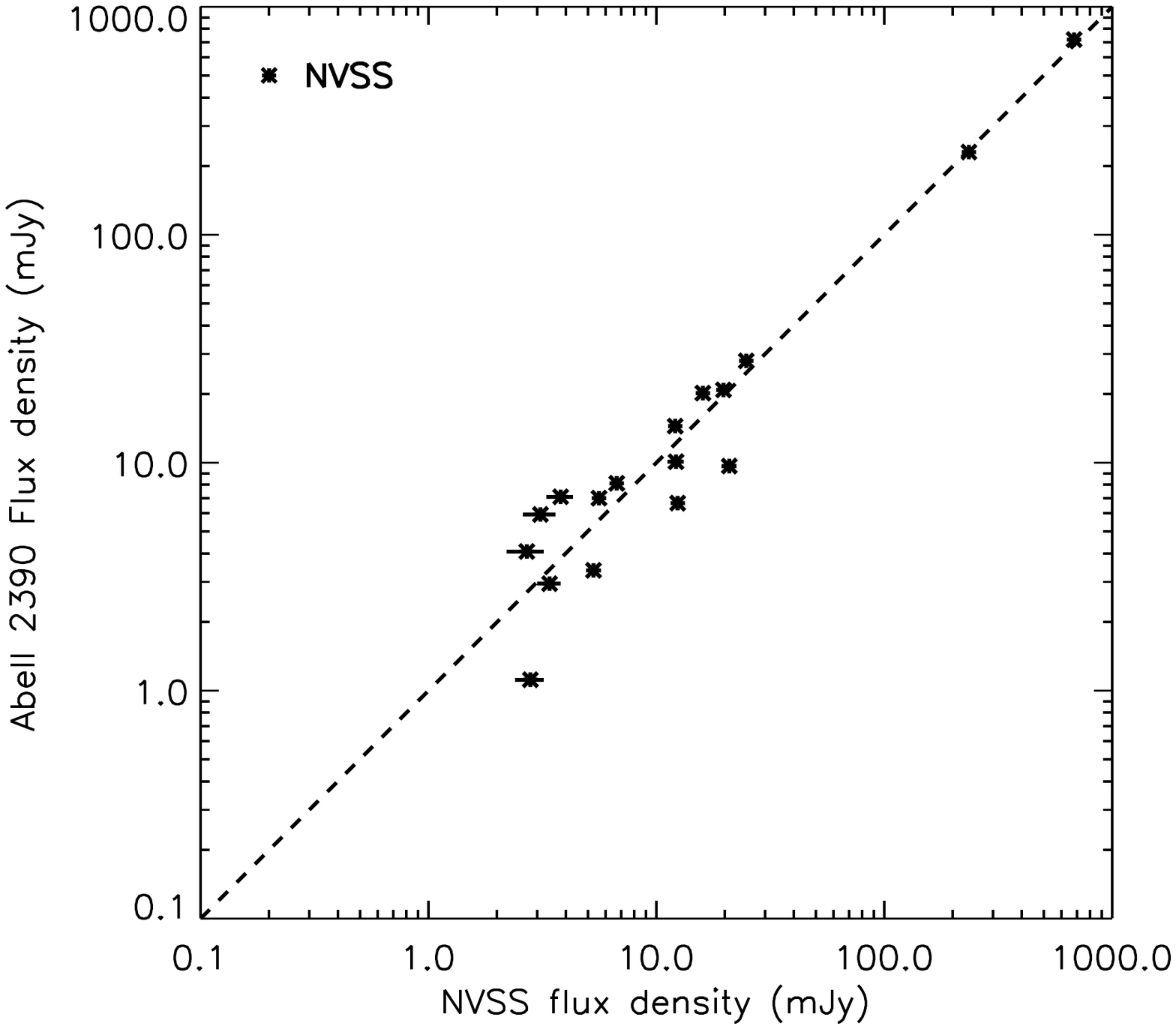}\caption{A2390 flux densities compared to those from the NVSS catalog. Of the
18 NVSS objects in our survey area, 17 (black asterisks) are found
to have counterparts. The dashed line indicates the ideal case, equality
of flux densities. }
\end{figure}
We find no systematic offset between our flux densities and the FIRST
and NVSS flux densities.

The FIRST survey does not cover the A2390 cluster field. The NVSS
catalog contains 18 objects within our A2390 field. Seventeen of these
objects have counterparts in our catalog. Based on the 16 sources
brighter than the $\sim$2.5 mJy NVSS sensitivity limit that also
have counterparts, the mean($S_{1.4\:{\rm GHz}}^{{\rm NVSS}}$/$S_{1.4\:{\rm GHz}}^{{\rm A2390}}$)
= 1.05$\pm$0.12. Figure 5 compares our flux density measurements
with those of NVSS. We find no systematic offset between our flux
densities and the NVSS flux densities.

We believe that most, if not all, non-overlapping NVSS and FIRST sources
are false detections. Inspection of our images reveals no viable counterparts.
The two NVSS sources without counterparts have cataloged flux densities
of 2.7 mJy (A370 field) and 2.8 mJy (A2390 field), which are very
close to the NVSS sensitivity limit. The non-overlapping object in
the A370 field also has no FIRST counterpart. The NVSS catalog only
gives upper limits for the major and minor deconvolved sizes, indicating
that these objects should be relatively compact. The nine non-overlapping
FIRST objects may be judged based on their reported sidelobe probability
values, which give the likelihood that a source is spurious. Spurious
sources are commonly due to artifacts created by nearby bright sources
\citep[][]{white08}. We find that there is a 60\% mean probability
that these 9 sources are spurious. Approximately 95\% of FIRST sources
have sidelobe probabilities that are lower than this mean value.

\section{Source Counts}

\begin{figure}
\centering{}\includegraphics[bb=180bp 80bp 660bp 540bp,clip,angle=180,scale=0.4]{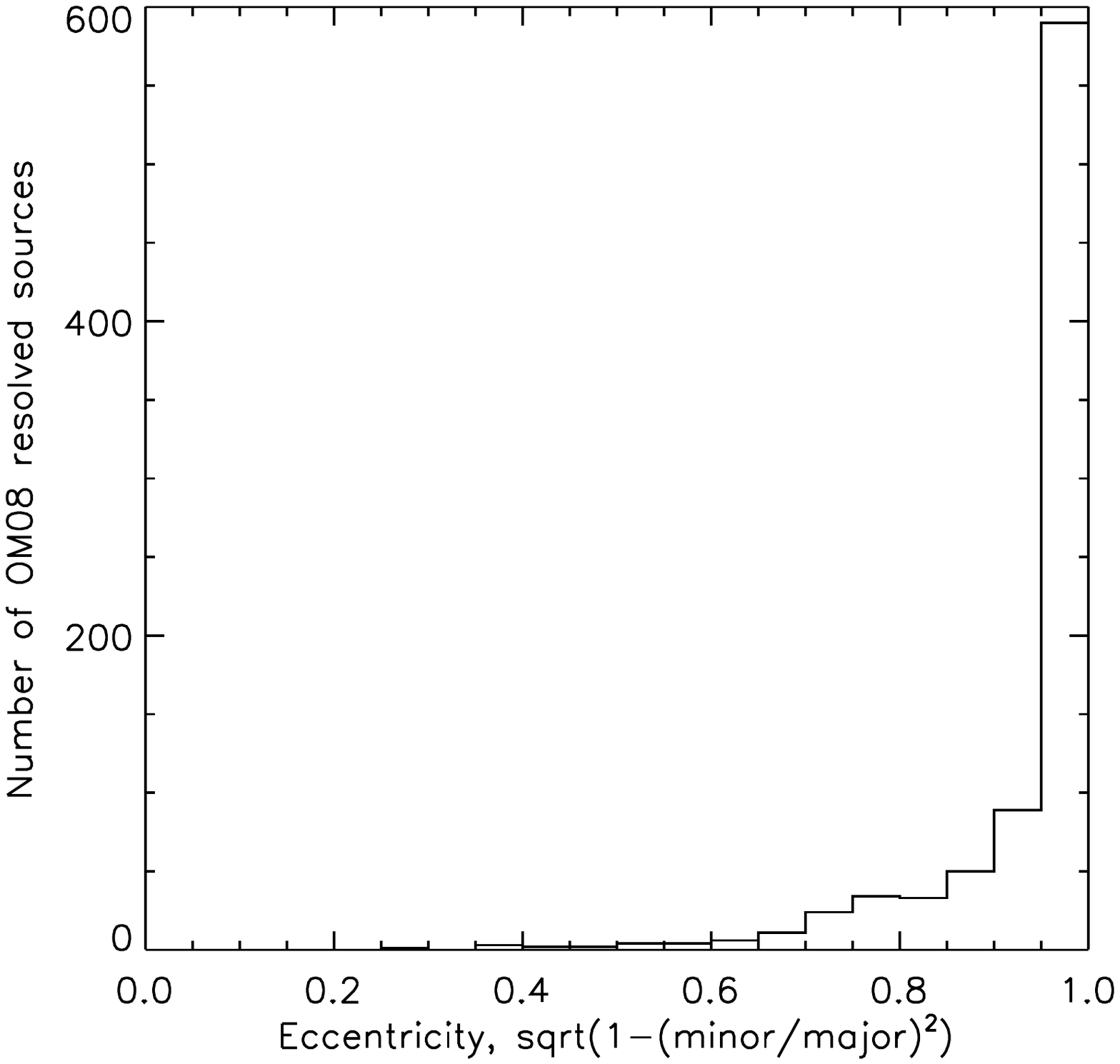}\caption{Eccentricity distribution of resolved OM08 radio sources. The high
level of observed eccentricity is the motivation behind our one-dimensional
simulated sources. These sources modified by instrumental effects
are used to assess the completeness of our catalog. }
\end{figure}
 
\begin{figure}
\centering{}\includegraphics[bb=0bp 15bp 985bp 1000bp,clip,scale=0.18]{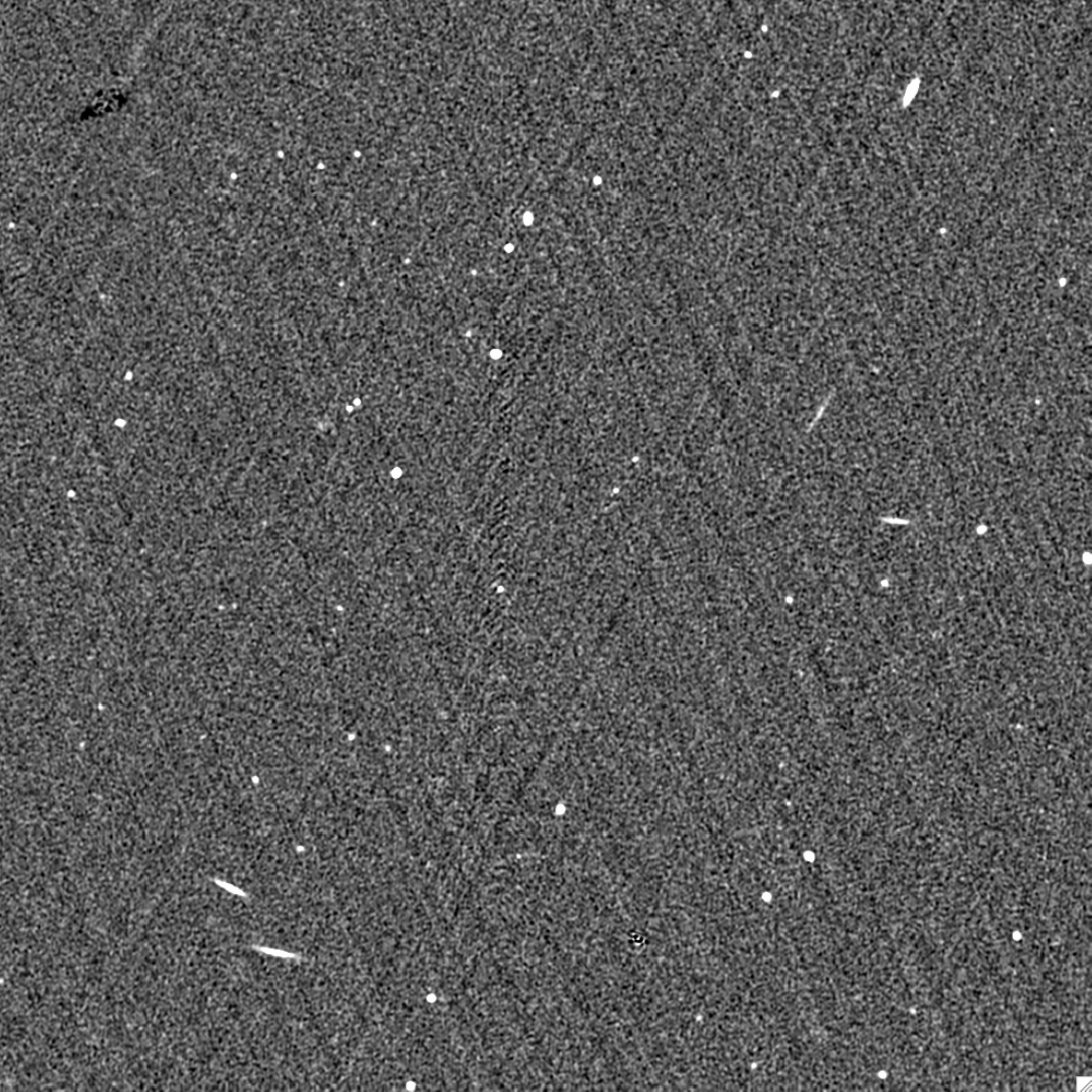}\caption{8$'$ $\times$ 8$'$ cutout of a simulated image. The recovery rate
of the simulated sources determines the correction factors applied
to our number counts. The background template is a residual, source
extracted, SAD image of the cluster field. SAD poorly extracts extended
sources, leaving a non-representative background. We prevent simulated
sources from populating these regions. The area affected is less than
1\% of the total field area. We show a region with a poorly extracted
extended source in the upper left corner. See Section 5 for details.}
\end{figure}
We constructed the differential source counts for the central region
(radius$<$16$'$) of each fields. These areas allow for the 5$\sigma$
detection of all point sources above $\sim$60 $\mu$Jy (see Figure
1). The A370 and A2390 catalogs contain 529 and 380 objects, respectively,
with 37 $<$ S$_{1.4{\rm \: GHz}}$ $<$ 3000 $\mu$Jy that lie within
the areas of interest. In the first three columns of Tables 3 and
4, we list the the 1.4 GHz flux bins, the flux bin centers in log
space, and the number of sources found per bin.

We assess the significance of false detections (5$\sigma$ noise spikes)
by performing an identical source extraction procedure on the inverted
radio image (see \citealt{ibar09} for a similar procedure). We generate
the inverted image by multiplying all pixel values by -1. This analysis
assumes that positive noise spikes are as likely as negative noise
spikes. Overall, we find that $\sim$8\% of our cataloged sources
are likely to be false detections (see Column 4 of Tables 3 and 4
for the number of false sources found per flux bin). We subtract the
false detections per bin from our raw counts. We show the small effect
of this correction on our final results in Figure 11.

To determine the completeness of our catalog, we developed a simple
Monte Carlo simulation that accounts for major instrumental and extraction
inefficiencies. For the background template, we used a residual, source
extracted, SAD image (see \citealt{biggs06} for a similar setup).
We removed all detections above $4.5$$\sigma$ from this background
image. We randomly positioned one thousand sources within the area
of interest, a 16$'$ radius circle. However, we prevented sources
from populating areas with poorly extracted extended sources. This
affected less than 1\% of the total area. We randomly sampled the
object's flux from a Euclidean power law distribution with a minimum
flux threshold of $4.5$$\sigma$. We randomly selected the object's
major axis from the OM08 distribution, shown in their Figure 8 and
assign a random position angle. Based on the observed eccentricity
of faint radio sources (e.g.\ see our Figure 6 for the OM08 eccentricity
distribution), we assumed an unresolved minor axis. All sources initially
had a Gaussian flux profile. We convolved the object with the beam
and then radially convolved the object with a Gaussian, with FWHM=(distance
from field center)$\times$$ $(channel bandwidth)$/$(central frequency).
The later convolution accounted for bandwidth smearing. Finally, we
modified the gain to account for primary beam attenuation. The simulation
did not account for finite time-average smearing, which we estimate
would to broaden sources by a few percent. We show an 8$'$$\times$8$'$
cutout of a simulated image in Figure 7.
\begin{figure}
\begin{centering}
\includegraphics[bb=190bp 80bp 660bp 540bp,clip,angle=180,scale=0.4]{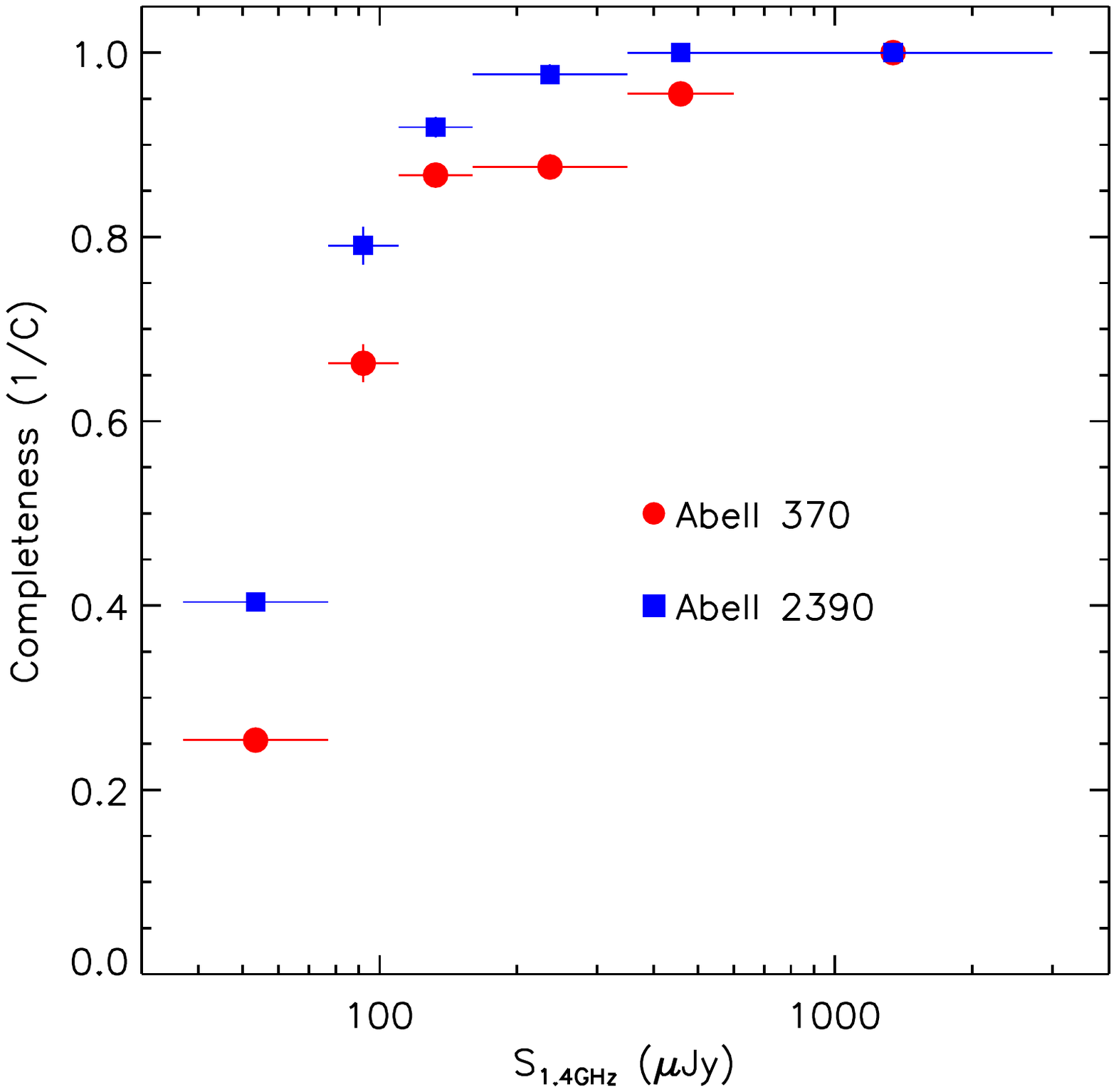}\caption{Monte Carlo simulation results. The completeness, or number of recovered
sources per number of input sources, as a function of flux density. }

\par\end{centering}

{\footnotesize (A color version of this figure is available in the
online journal.)}
\end{figure}
\begin{figure}
\begin{centering}
\includegraphics[bb=190bp 80bp 660bp 540bp,clip,angle=180,scale=0.4]{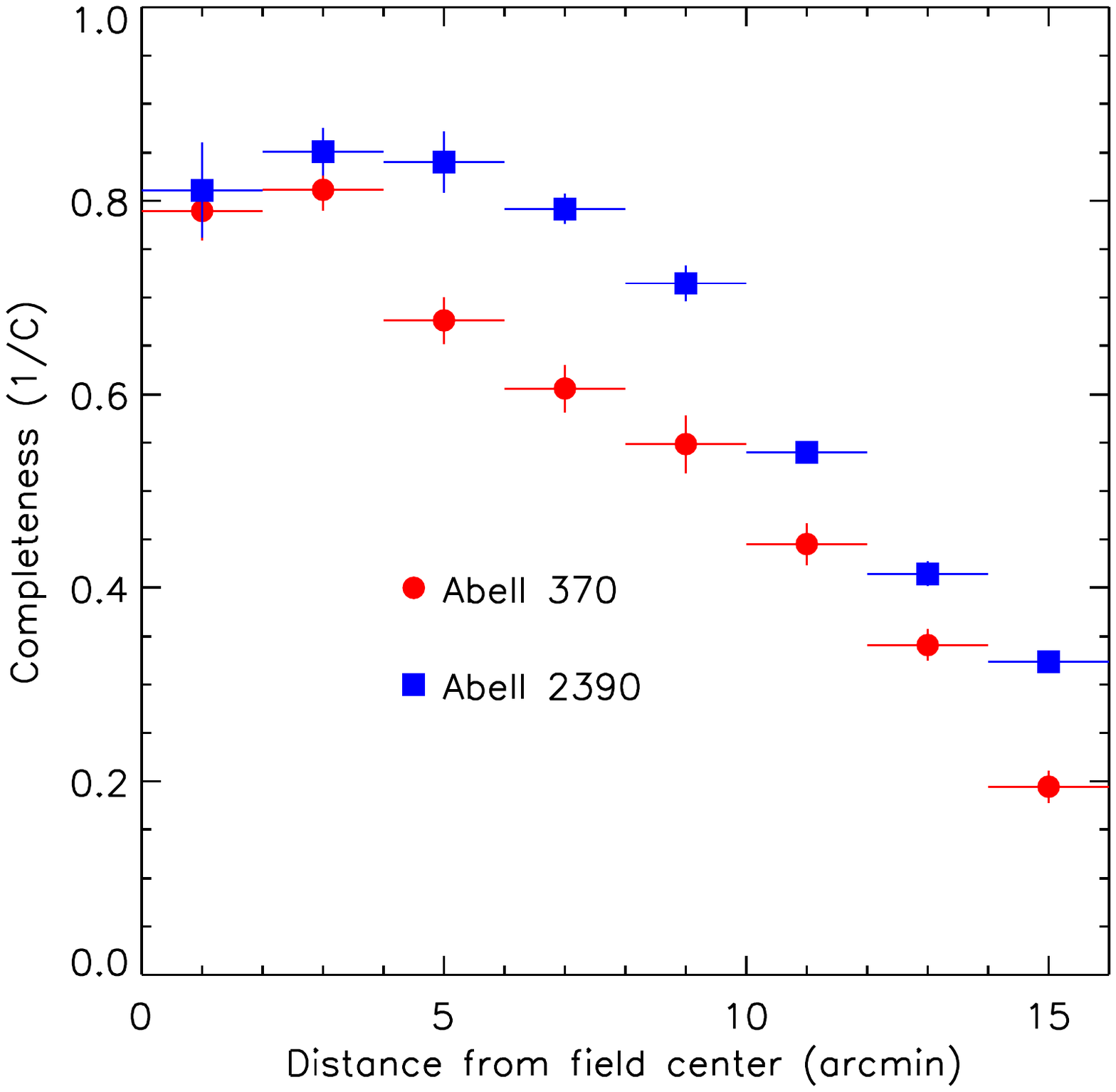}\caption{Monte Carlo simulation results. The completeness, or number of recovered
sources per number of input sources, as a function of radius from
field center. }

\par\end{centering}

{\footnotesize (A color version of this figure is available in the
online journal.)}
\end{figure}
\begin{figure*}[!t]
\begin{centering}
\includegraphics[bb=55bp 55bp 710bp 530bp,clip,angle=180,scale=0.5]{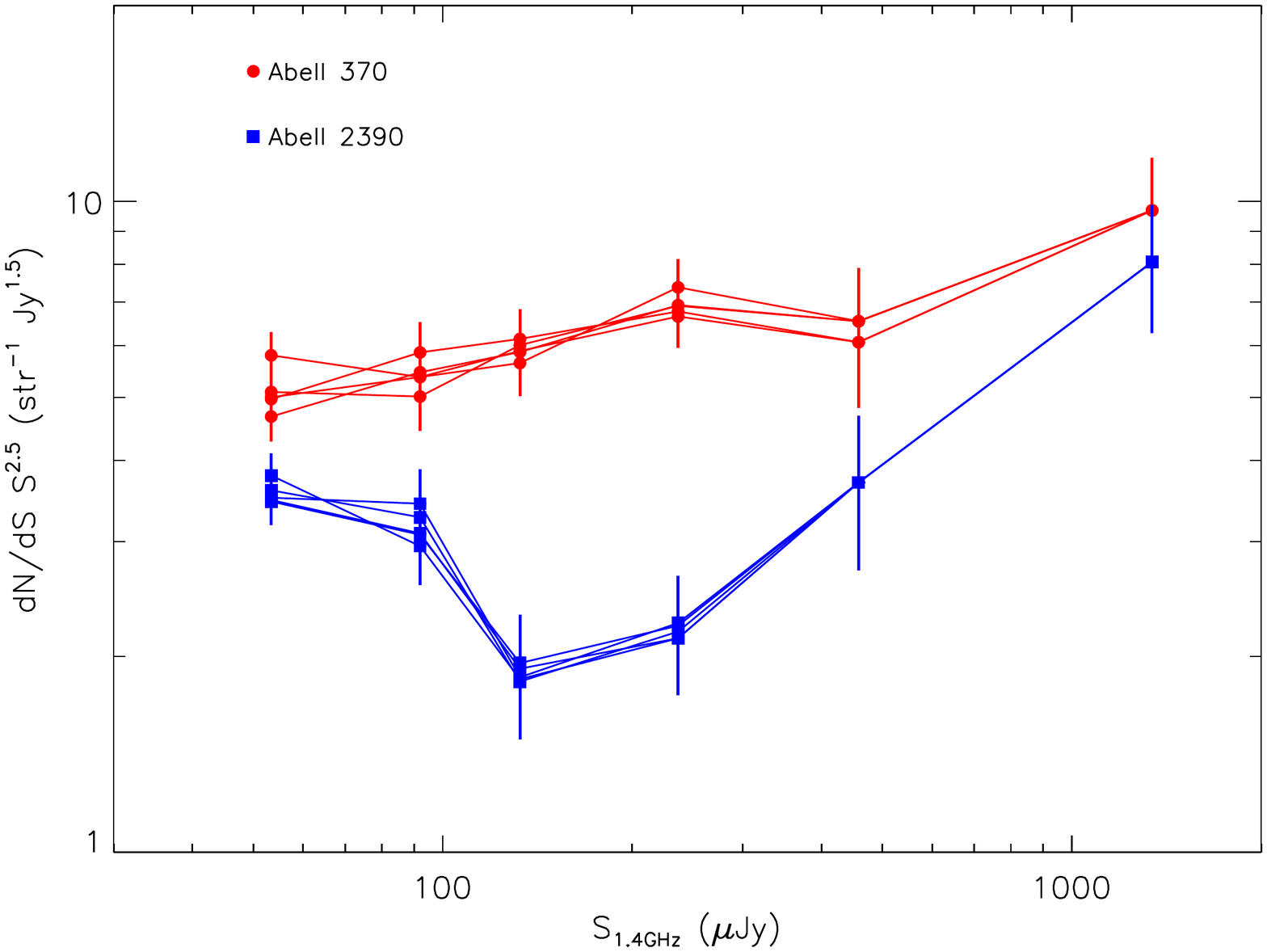}\caption{Five Monte Carlo completeness results per field applied to our A370
and A2390 counts. We use the variation of the Monte Carlo results
to estimate the error in the derived correction factors. This error
is less than or approximately equivalent to the Poisson error (displayed
error bars). }

\par\end{centering}

{\footnotesize (A color version of this figure is available in the
online journal.)}
\end{figure*}
\begin{figure*}[!t]
\begin{centering}
\includegraphics[bb=55bp 55bp 710bp 530bp,clip,angle=180,scale=0.5]{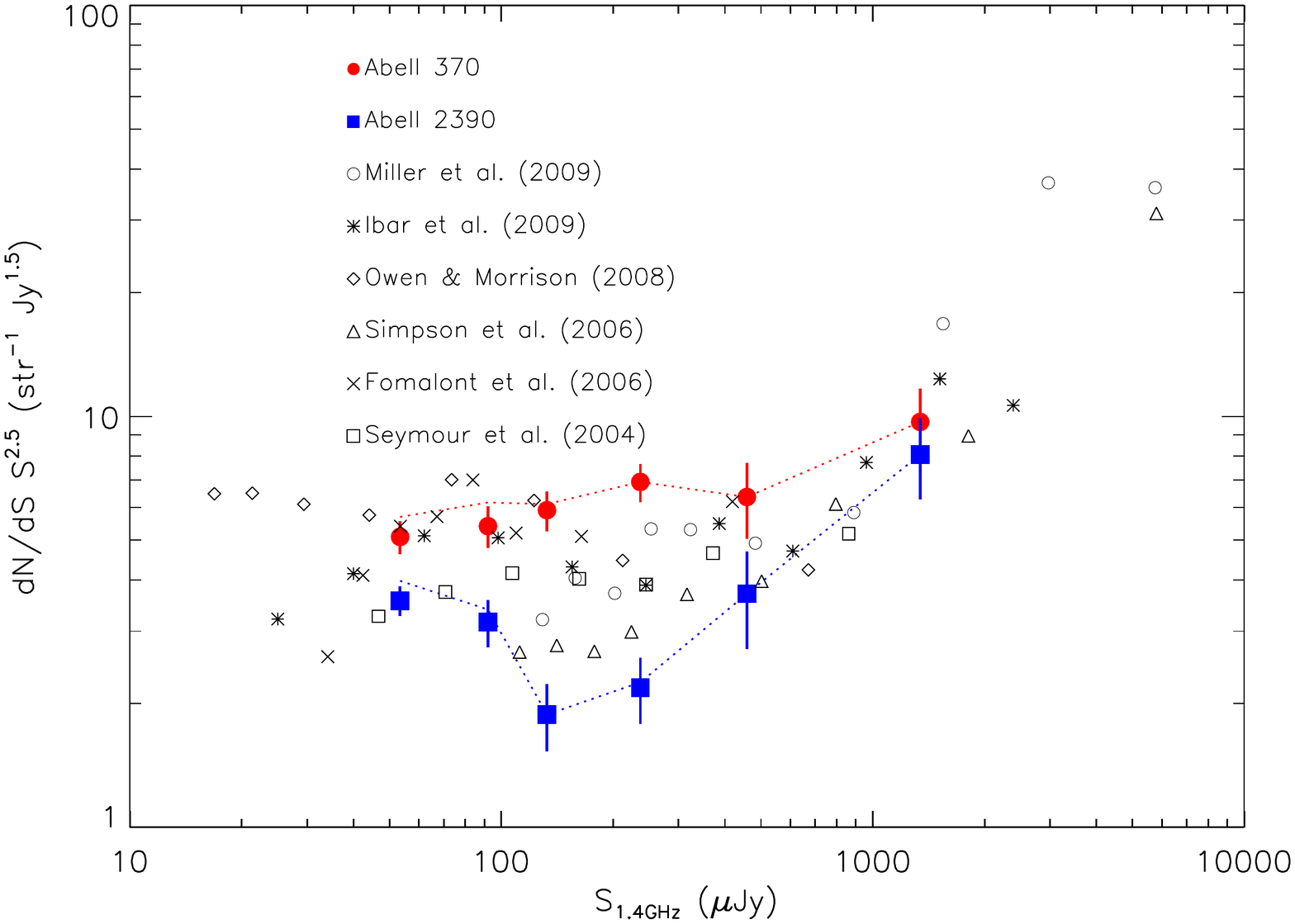}\caption{A370 and A2390 differential source counts compared to blank field
counts and Coma cluster counts \citep{miller09}. The dotted lines
indicate the effect on our derived counts of not correcting for false
detections.}

\par\end{centering}

{\footnotesize (A color version of this figure is available in the
online journal.)}
\end{figure*}
\begin{table}[!h]
\caption{A370 1.4 GHz radio source counts.}

\centering{}%
\begin{longtable}{cccccc}
\hline
\hline 
\begin{tabular}{c}
S bin\tabularnewline
($\mu$Jy)\tabularnewline
\end{tabular} & %
\begin{tabular}{c}
S\tabularnewline
($\mu$Jy)\tabularnewline
\end{tabular} & N & N$_{false}$ & C & $ $%
\begin{tabular}{c}
S$^{2.5}$dN/dS\tabularnewline
(Jy$^{1.5}$sr$^{-1}$)\tabularnewline
\end{tabular}\tabularnewline
\hline 
\endhead
37 - 77 & 53 & 189 & 20 & 3.9 & 5.1$\pm$0.5\tabularnewline
77 - 110 & 92 & 113 & 14 & 1.5 & 5.4$\pm$0.6\tabularnewline
110 - 160 & 133 & 89 & 3 & 1.2 & 5.9$\pm$0.7\tabularnewline
160 - 350 & 237 & 91 & 0 & 1.1 & 6.9$\pm$0.7\tabularnewline
350 - 600 & 458 & 23 & 0 & 1.0 & 6.4$\pm$1.3\tabularnewline
600 - 3000 & 1342 & 24 & 0 & 1.0 & 9.7$\pm$2.0\tabularnewline
\end{longtable}
\end{table}
\addtocounter{table}{-1}
\begin{table}[!h]
\caption{A2390 1.4 GHz radio source counts.}

\centering{}%
\begin{longtable}{cccccc}
\hline
\hline 
\begin{tabular}{c}
S bin\tabularnewline
($\mu$Jy)\tabularnewline
\end{tabular} & %
\begin{tabular}{c}
S\tabularnewline
($\mu$Jy)\tabularnewline
\end{tabular} & N & N$_{false}$ & C & %
\begin{tabular}{c}
S$^{2.5}$dN/dS\tabularnewline
(Jy$^{1.5}$sr$^{-1}$)\tabularnewline
\end{tabular}\tabularnewline
\hline 
\endhead
37 - 77 & 53 & 210 & 22 & 2.5 & 3.6$\pm$0.3\tabularnewline
77 - 110 & 92 & 74 & 5 & 1.3 & 3.2$\pm$0.4\tabularnewline
110 - 160 & 133 & 29 & 0 & 1.1 & 1.9$\pm$0.3\tabularnewline
160 - 350 & 237 & 33 & 1 & 1.0 & 2.2$\pm$0.4\tabularnewline
350 - 600 & 458 & 14 & 0 & 1.0 & 3.7$\pm$1.0\tabularnewline
600 - 3000 & 1342 & 20 & 0 & 1.0 & 8.1$\pm$1.8\tabularnewline
\end{longtable}
\end{table}
\addtocounter{table}{-1} 

We perform the extraction technique described in Section 3 on the
simulated image. We complete this procedure 5 times, providing a total
of 5000 simulated sources per field. The recovery rate between the
known input and the extracted output gives the correction factors,
$C$, listed in Column 5 of Tables 3 and 4 and displayed in Figures
8 and 9. In Figure 10, we show how the variation of the 5 Monte Carlo
simulations affects our derived counts. We calculate the corrected
counts using the following equation: 
\begin{figure*}[!t]
\centering{}\includegraphics[bb=55bp 55bp 710bp 530bp,clip,angle=180,scale=0.5]{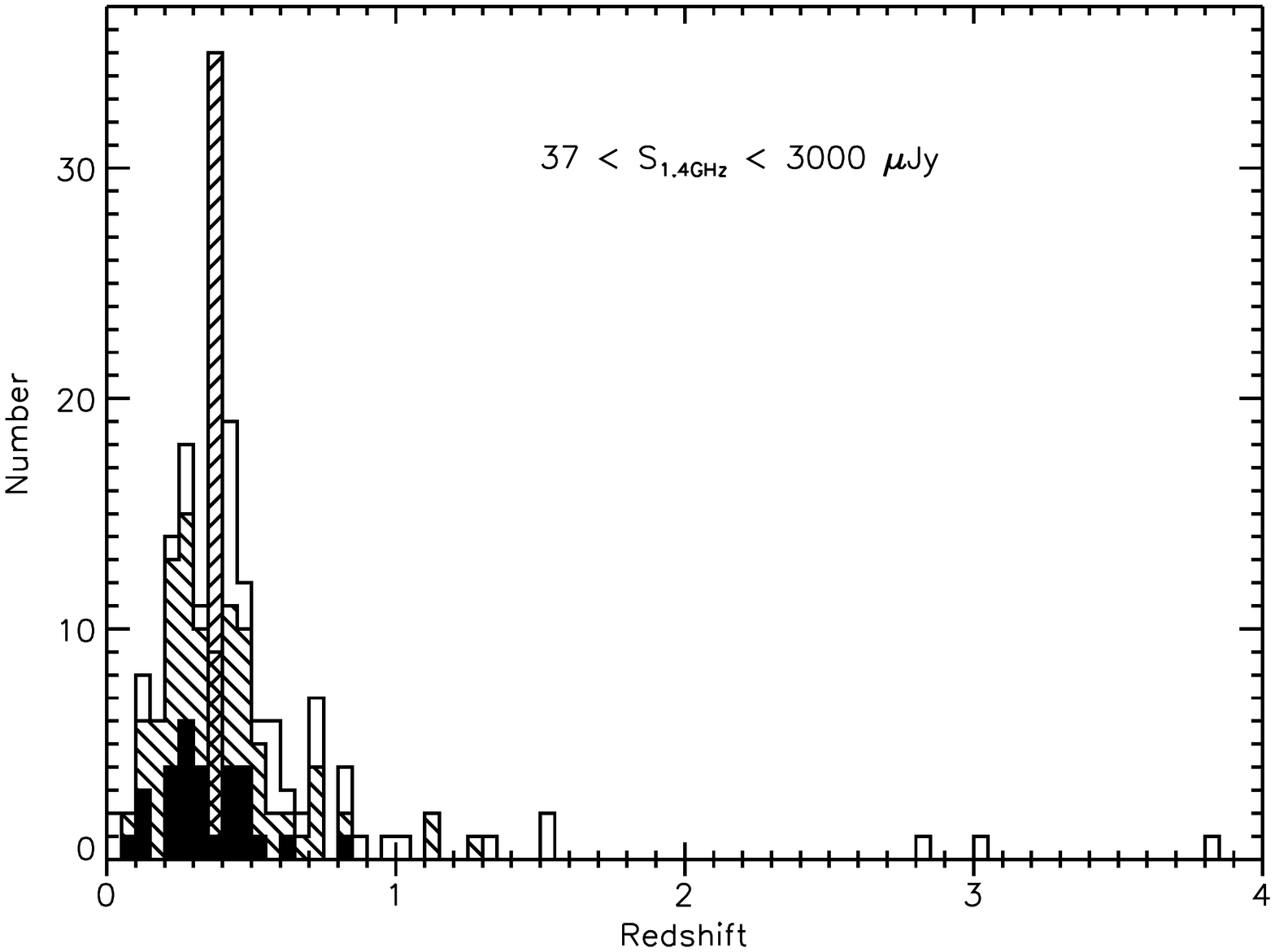}\caption{A histogram of the 167 redshifts in our A370 number counts sample.
The outer outline of the histogram indicates the redshift distribution
of the entire 167 objects. The three interior filled histograms display
the effect of applying cluster masks. The histogram filled with positive-slope
diagonal lines designates the 35 cluster members that were excluded
in our $z$$_{A370}\pm$0.025 mask. The histogram filled with negative-slope
diagonal lines designates the spectroscopically identified sources
that lie beyond a 6$'$ radius from the cluster center. Our 6$'$
radius cluster mask excludes 26 (74\%) of the spectroscopically identified
cluster members. The filled histogram designates the spectroscopically
identified sources that lie beyond 12$'$ radius from the cluster
center. Our 12$'$ radius cluster mask excludes 34 (97\%) of the spectroscopically
identified cluster members.}
\end{figure*}

\[
N_{corr}=(N-N_{false})\times C\,.
\]

\noindent We then divide $N{}_{corr}$ by the survey area and the
bin width. We normalize these values by the Euclidean slope $S^{-2.5}$,
where $S$ is bin center in log space. We summarize the final results
in Column 6 of Tables 3 and 4. In Figure 11, we show the A370 (\textit{red
circles}) and A2390 (\textit{blue squares}) number counts with error
bars indicative of Poisson uncertainty and Monte Carlo variation.
We use dotted lines to indicate the effect on our derived counts of
not correcting for false detections. We use open black circles to
show number counts constructed from a VLA Coma cluster survey \citep*{miller09}.
We also show various blank field counts (\textit{black: asterisk}s,
\textit{diamond}s, \textit{triangle}s, \textit{cross}es, and \textit{square}s)
for comparison to our A370 and A2390 number counts.

\section{Discussion }

The A370 number counts 
\begin{figure*}[!t]
\begin{centering}
\includegraphics[bb=55bp 55bp 710bp 530bp,clip,angle=180,scale=0.5]{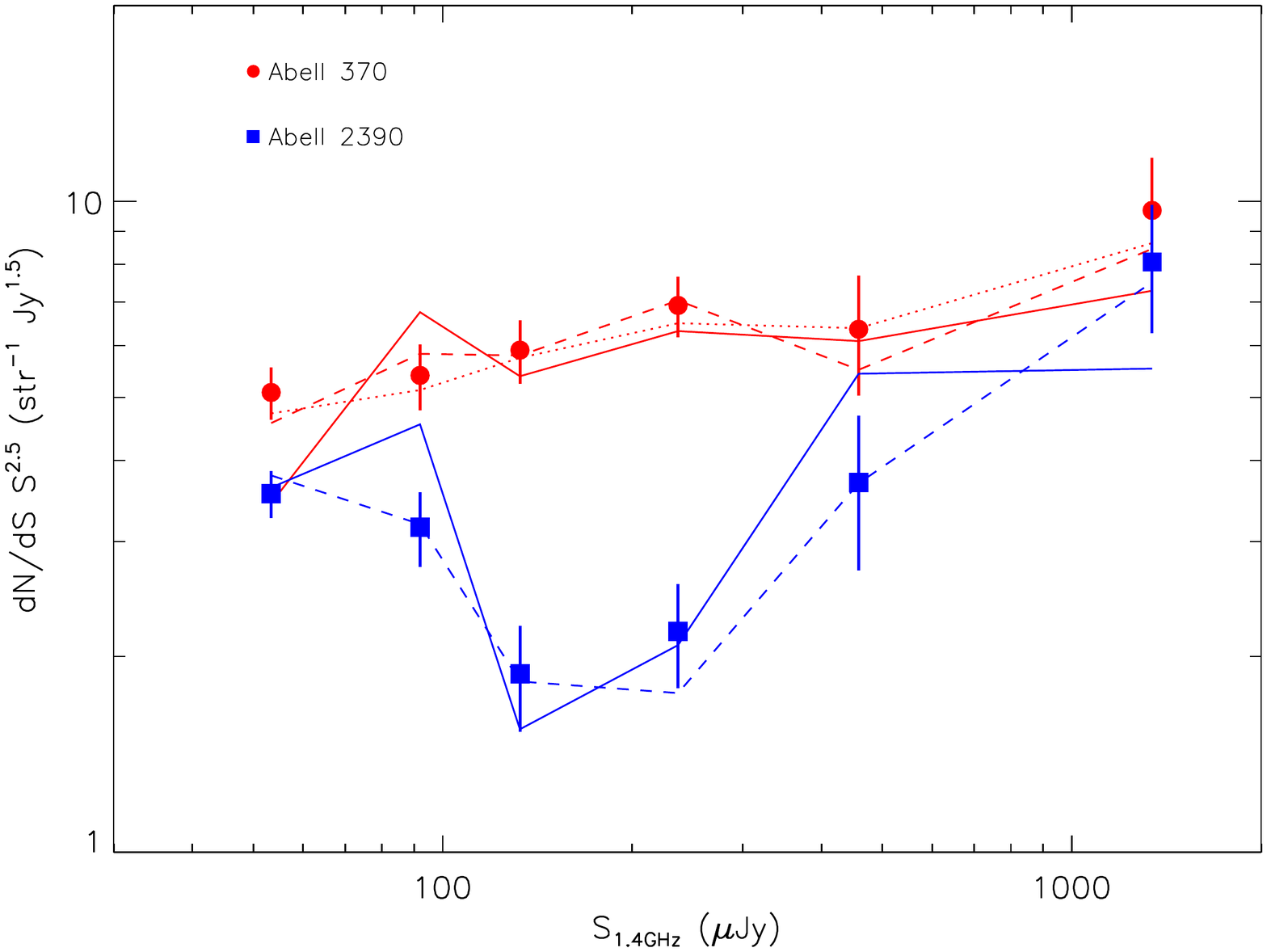}\caption{Illustration of the effect on our derived number counts of applying
radial cluster masks. Symbols with error bars indicate the cluster
counts with no mask applied (full 16$'$ radius field). The dashed
lines indicate the counts with a radial area of 6$'$ from the cluster
centers masked. The solid lines indicate the counts with a radial
area of 12$'$ from the cluster centers masked. The red dotted line
indicates the counts with spectroscopically confirmed A370 cluster
members excluded. On average, the A370 counts are reduced, bringing
them in closer agreement to blank field results. The A2390 counts
are consistently low. Even taking into account the greater Poisson
error of the more restrictive masks, the counts in the two cluster
fields are found to be inconsistent with one another in the 110 -
350 $\mu$Jy flux range.}

\par\end{centering}

{\footnotesize (A color version of this figure is available in the
online journal.)}
\end{figure*}
form an upper envelope to the blank field number counts. This
could potentially be explained by an over-density of cluster galaxies
superimposed on the $z$ $\sim$ 1 field galaxies. In addition to
the galaxy cluster at $z$ = 0.375, Keenan et al. (2012) found galaxy
over-densities at $z$ $\sim$ 0.18 and $z$ $\sim$ 0.25. We clearly
see evidence for the $z$ $\sim$ 0.25 over-density in the redshift
distribution (see Figure 12). $ $

The A2390 number counts form a lower envelope to the blank field number
counts, which is more surprising. This might indicate that the galaxy
cluster at $z$=0.228 does not significantly contribute to a sparsely
populated field of 37 $<$ S$_{1.4\:{\rm GHz}}$ $<$ 3000 $\mu$Jy
galaxies. 

In the following, we investigate disentangling the cluster galaxy
population from the field galaxy population using two techniques.
1) \textit{Redshift mask}: For the A370 field, we re-derive the number
counts with all the spectroscopically confirmed cluster members excluded.
We lack sufficient spectroscopic data to perform a similar analysis
on A2390. 2) \textit{Radial masks}: We re-derive the number counts
for both fields with the cluster center masked. In other words, we
re-derive the number counts within a range of annuli to eliminate
the majority of cluster objects. As we reduce the influence of the
cluster, we note the effect that has on the number counts.

\subsection{Redshift Mask}

Our large sample of A370 redshifts allows us to re-derive the number
counts after excluding known cluster members. To construct the A370
number counts, we used the 529 sources with 37 $<$ S$_{1.4{\rm \: GHz}}$
$<$ 3000 $\mu$Jy that lie within 16$'$ of the field center. Of
these, 167 (32\%) have known redshifts. In Figure 12, we show the
distribution of these redshifts. We define cluster members as any
object with $z$$_{A370}\pm$0.025. Rather than a physical cluster
scale, we base the redshift range on a conservative estimate of the
redshift error. Using this criteria, we identify 35 objects as cluster
members, and we exclude them from the re-derived counts. In Figure
12, we use positive-slope diagonal lines to denote the cluster members.
We note that our spectroscopic completeness declines as the 1.4 GHz
flux decreases and as the projected distance from the cluster center
increases. Given these biases and the proximity of the cluster members
compared to the average field galaxy at $\left\langle {\rm z}\right\rangle $
$\sim$ 1, we expect to have succeeded in excluding more than 32\%
of the cluster members. 

We show the re-derived number counts in Figure 13 as a red dotted
line. We find that the A370 counts are lowered by a weighted average
of 5.7\%$\pm$6.5\%. The estimated error in our weighted average calculation
(6.5\%) results from propagating the Poisson error of our measured
counts. Although not corrected for, excluding spectroscopic cluster
members will also remove the cluster's effective volume. This will
reduce the survey's effective area and therefore increase the re-derived
counts. This effect is small (estimated to be on the order of a few
percent), since the effective volume of our radio field is large when
compared to the effective volume of the cluster. Given our spectroscopic
completeness of 32\%, we expect the systematic uncertainty in our
estimate of cluster influence to be dominated by any unidentified
cluster members. Thus, we regard the reported 5.7\%$\pm$6.5\% reduction
in our re-derived counts as a lower limit.

\subsection{Radial Masks}

We also re-derive the number counts with cluster center radial masks.
We apply masks to eliminate all galaxies within 2$'$, 4$'$, 6$'$,
8$'$, 10$'$, and 12$'$ from the cluster center. We obtain an estimate
of the effectiveness of removing cluster members from both fields
using radial masks by considering the cluster evolution study of \citet[][hereafter M99]{morrison99}.
M99 investigated the radial distribution of low-luminosity (10$^{22.3}$$<$
L$_{1.4\:{\rm GHz}}$ $<$ 10$^{23}$ W Hz$^{-1}$) radio galaxies
(LLRGs) for a $z$ $<$ 0.25 and a $z$ $\sim$ 0.4 cluster sample.
The low-redshift sample was composed of 76 cluster LLRGs, while the
z$\sim$0.4 sample was composed of 43 cluster LLRGs. 

From M99's derived cumulative radial density profile (his Figure 6.18),
we see that a 1.9 Mpc (6$'$ radius) A370 mask should exclude $\sim$82\%
of the cluster's LLRG population, while a 3.7 Mpc (12$'$ radius)
mask should exclude $>$97\%. The 1.3 Mpc (6$'$) and 2.6 Mpc (12$'$)
A2390 masks should exclude $\sim$82\% and $\sim$97\% of the cluster's
LLRGs, respectively. LLRGs correspond to galaxies with flux densities
between 41 - 206 $\mu$Jy for the A370 cluster and 129 - 648 $\mu$Jy
for the A2390 cluster. The radial distributions of higher luminosity
radio sources are found to be more centrally compact. Thus, we expect
our masks to reduce significantly the population of all S$_{1.4\:{\rm GHz}}>$129
$\mu$Jy ($>$41 $\mu$Jy for A370) cluster sources. There are uncertainties
in this estimate, since individual clusters will have deviations from
this averaged radial profile. Additionally, the M99 study notes a
lack of confirmed outer cluster members in the $ $$z$ $\sim$ 0.4
sample, which may bias this radial profile. As a check on these percentages,
we note how many spectroscopically confirmed cluster members remain
after applying the A370 radial masks. Figure 12 shows the redshift
histogram before and after the application of the 6$'$ (histogram
filled with negative-slope diagonal lines) and the 12$'$ (filled
histogram) masks. Our 6$'$ and 12$'$ cluster masks remove 74\% and
97\% of the spectroscopically identified cluster members, respectively.
This estimate, while roughly consistent with the M99 result, suffers
from our spectroscopic incompleteness, which gets worse with increasing
radius and decreasing flux. We conclude that our 6$'$ radial masks
should significantly reduce cluster influence on the derived number
counts. We adopt the 12$'$ mask result as our best estimate of the
total cluster influence. 

In Figure 13, we show the number counts after excluding radial areas
of 6$'$ (\textit{dashed}) and 12$'$ (\textit{solid}) from the cluster
centers. These angular radii, respectively, correspond to 1.9 and
3.7 Mpc at $z$$_{A370}$ = 0.375 and 1.3 and 2.6 Mpc at $z$$_{A2390}$
= 0.228. The A370 field has a 5.0 Mpc (16$'$) radial extent, while
the A2390 field has a 3.5 Mpc (16$'$) radial extent. Excluding radial
areas of 6$'$ and 12$'$ from the A370 cluster center reduces the
number counts by a weighted average of 3.5\%$\pm$7.4\% and 7.6\%$\pm$10.7\%.
If we also exclude any additional spectroscopically identified cluster
members not in the masked regions, then these percentages increase
by 1\%. Excluding radial areas of 6$'$ and 12$'$ from the A2390
cluster center reduces the number counts by an average of 0.7\%$\pm$8.8\%
and 1.1\%$\pm$15.1\%. In Figure 13, we can see that the re-derived
number counts in A2390 are consistently low in the 110 to 350 $\mu$Jy
range, even when compared to the lowest blank field results. 

Our best estimate indicates that the A370 cluster influences the derived
number counts on the 10$\%$ level. The A2390 cluster does not appear
to alter significantly the derived number counts. However, the large
Poisson error of our measurements weakens this conclusion.

\subsection{Comparison to Coma Cluster Counts}

Coma is a local ($z$=0.0231) cluster of galaxies, while our clusters
are at redshifts of $z$=0.228 and $z$=0.375. In cluster environments,
the fraction of LLRGs decreases with decreasing redshift (M99). Thus,
evolution effects may be significant and no direct comparison is possible.
However, we may compare our results with the Coma cluster counts merely
to check for consistency given this known evolution with redshift.
Specifically, our finding that $z$$\sim$0.3 cluster LLRGs are not
a major component in the derived number counts would be at odds with
a local measurement showing a significant LLRG contribution.

\citet[][]{miller09} surveyed two $\sim$0.5 deg$^{2}$ 1.4 GHz VLA
observations covering the core and southwest region of the Coma cluster.
They presented number counts from 0.110 mJy (their 5$\sigma$ limit)
to 100 $ $mJy. Their counts have only been corrected for areal coverage.
Thus, their faint counts ($\lesssim$ 0.230 mJy) should be regarded
as lower limits. At the Coma cluster's redshift LLRGs would correspond
to the flux density range 16-82 mJy. No overdensity is noted in this
range. Moreover, they find that their cluster counts are consistent
with blank field counts (we also show this in Figure 11). The only
excess of sources is in the $\sim$1-5 mJy range, and this is not
attributed to a cluster effect. We conclude that our result, indicating
that field galaxies make up the vast majority of sources in our cluster
fields, is consistent with the local counts found in the Coma cluster
field.

\subsection{Gravitational Lensing Bias}

Gravitational lensing affects number counts by boosting observed source
flux and by magnifying the source plane. This lensing bias occurs
over a relatively small area compared to our r $=$ 16$'$ fields.
Given reasonable estimates for the cluster masses and background source
redshifts, we estimate that beyond 2 Mpc there is no significant lensing
bias. We believe our estimate of cluster influence does not require
any modifications due to lensing considerations.

\subsection{Cosmic Variance}

Our results are consistent with the faint number counts being primarily
determined by field radio galaxies. We performed similar reduction
and extraction procedures on both cluster fields. Further, the lack
of A2390 sources is already seen in the bright radio flux bins, and
these should not suffer significantly from incompleteness. For these
reasons, we investigate the significance of cosmic variance.

We estimate the effect of cosmic variance with the method of \citet{somerville04}.
Assuming a correlation length of 5 Mpc (consistent with the values
given by \citet{overzier03} for NVSS mJy sources) and a redshift
interval of $z$ = 1 $\pm$ 0.5, we estimate $\sim$12\% rms variance
due to large-scale structure (see \citealt{simpson06} for a similar
calculation). 

Since there is no clear consensus on the `true' faint number counts,
we choose to determine if our derived cluster number counts for the
two fields can be brought into agreement given reasonable assumptions.
This avoids having to develop some ad hoc method of averaging together
faint number counts from radio surveys that have different analysis
methods. An accurate comparison of our derived number counts depends
on the accuracy of our derived error bars. While the estimated error
does take into account Poisson error and the variance of our Monte
Carlo simulations, it does not account for uncertainties inherent
to the completeness simulations. For example, we determined a simulated
object's major axis by randomly sampling from OM08's size distribution.
Any inaccuracy in this distribution could alter our derived completeness
corrections and thus, our derived number counts. We adopt 1.5$\sigma$
error bars to account for these additional concerns. Assuming a 10\%
reduction in the A370 number counts to account for the cluster's influence,
we find that our counts can be brought to within 1 - 2 sigma of each
other. This suggests that cosmic variance can explain the disagreement
seen in our derived number counts.

\section{Summary}

We presented 1.4 GHz catalogs for the cluster fields A370 and A2390
observed with the Very Large Array. These are two of the deepest radio
images of cluster fields ever taken and will be useful for both cluster
and field studies. The A370 image covers an area of 40$'$ $\times$
40$'$ with a synthesized beam of $\sim$1.7$''$ and a noise level
of $\sim$5.7 $\mu$Jy near field center. The A2390 image covers an
area of $\sim$34$'$ $\times$ 34$'$ with a synthesized beam of
$\sim$1.4$''$ and a noise level of $\sim$5.6 $\mu$Jy near field
center. We cataloged 200 redshifts for the A370 field. For the region
within 16$'$ of field center, we derived differential number counts
for both fields. We found that the faint (37 $<$ S$_{1.4\:{\rm GHz}}$
$<$ 3000 $\mu$Jy) number counts of A370 are roughly consistent with
the highest blank field counts, while the faint number counts of A2390
are roughly consistent with the lowest blank field counts. For the
A370 field, we found that cluster members increase our derived number
counts by $\sim$10\%, while the A2390 cluster does not appear to
contribute significantly to the presented faint number counts. 

We suggest that the disagreement between our faint number counts can
be primarily attributed to cosmic variance. We note that our observed
fields, consisting of single pointings with nonuniform sensitivity
and contaminated by foreground clusters, are not ideal to resolve
definitively the debate over the importance of cosmic variance on
faint radio sources. Further progress awaits a large area survey that
probes the micro-Jansky population.

\acknowledgements{We would like to thank the referee for their critical reading of
the paper and useful suggestions for improving it. We gratefully acknowledge
support from NSF grant AST 0708793, the University of Wisconsin Research
Committee with funds granted by the Wisconsin Alumni Research Foundation
and the David and Lucile Packard Foundation (A. J. B.). Additional
support was provided by a Wisconsin Space Grant Consortium Graduate
Fellowship, a Sigma Xi Grant in Aid of Research, and a National Radio
Astronomy Observatory Graduate Internship (I. G. B. W.). The National
Radio Astronomy Observatory is a facility of the National Science
Foundation operated under cooperative agreement by Associated Universities,
Inc.}

\clearpage
\end{document}